\documentclass[
twocolumn,
pre,
a4paper,
nofootinbib,
amsmath,
floatfix
]{revtex4}

\let \Onecolumn

\newif\ifOnecolumn
  \ifx\Onecolumn\undefined
  \Onecolumntrue
\else
  \Onecolumnfalse
\fi
 
\usepackage{graphicx,amssymb,amsmath,amsfonts}
\usepackage[mathscr]{eucal}

\usepackage{comment}

\begin{document}

\title{Critical Casimir interaction of ellipsoidal colloids with a planar wall}

\author{S. Kondrat}  \author{L. Harnau}   \author{S. Dietrich}
\affiliation{Max-Planck-Institut f\"ur Metallforschung,  
Heisenbergstr.~3, D-70569 Stuttgart, Germany}
\affiliation{Institut f\"ur Theoretische und Angewandte Physik,
Universit\"at Stuttgart, Pfaffenwaldring 57, D-70569 Stuttgart, Germany}

\date{\today}

\begin{abstract}

Based on renormalization group concepts and explicit mean field
calculations we study the universal contribution to the effective force
and torque acting on an ellipsoidal colloidal particle which is
dissolved in a critical fluid and is close to a homogeneous planar
substrate. At the same closest distance between the substrate and the
surface of the particle, the ellipsoidal particle prefers an orientation
parallel to the substrate and the magnitude of the fluctuation induced 
force is larger than if the orientation of the particle is perpendicular 
to the substrate. The sign of the critical torque acting on the ellipsoidal
particle depends on the type of boundary conditions for the order
parameter at the particle and substrate surfaces, and on the pivot with
respect to which the particle rotates.

\end{abstract}

\maketitle

\section{Introduction}
\label{sec:intro}

The confinement of the order parameter fluctuations in a critical fluid
leads to an effective long-ranged interaction between the confining
walls and colloidal particles suspended in the fluid. The occurrence of
such a force was predicted by \citeauthor{fisher:78} in
\citeyear{fisher:78} \cite{fisher:78}. It is called critical (or
thermodynamic) Casimir force \cite{krech:book, danchev:book}, in analogy
with the Casimir force \cite{casimir:48} in quantum electrodynamics
where the force originates from the confined quantum fluctuations of the
electromagnetic field \cite{kardar:99:rmp}.

Since then the critical Casimir effect has attracted increasing
theoretical and experimental attention. So far the theoretical
investigations of the critical Casimir effect have been focused on the
film geometry, realized either by homogeneous, planar, and parallel
walls (see, e.g., Refs. \cite{krech:91:prl, krech:92:pra:1886,
krech:92:pra:1922, evan:94:prb:8842, krec:97, law:01, vasi:07,
maciolek:06, maciolek:07, dantchev:07:pre, huch:08,
grueneberg:08:prb:115409}  and references therein) or by chemically
patterned \cite{sprenger:06} or geometrically structured substrates
\cite{troe:08}, as well as on spherical colloidal particles (see, e.g.,
Refs. \cite{eisenriegler:95, burkhardt:95,hanke:prl:98,schlesener:03} 
and references therein). Strong experimental evidences for critical
Casimir forces have been obtained by studying wetting films near
critical end points of quantum \cite{garcia:99, garcia:02, ganshin:06}
or classical fluids \cite{fukuto:05, rafai:07}.

It has been pointed out by \citeauthor{fisher:78} \cite{fisher:78,
degennes:81} that such fluctuation-induced forces should lead to
flocculation of colloidal particles if their solvent is a binary liquid
mixture close to its consolute point. Such a solvent-mediated
flocculation, which can be interpreted as indirect evidence for the
critical Casimir force, has indeed been observed for silica spheres
suspended in a binary liquid mixture of water and lutidine
\cite{beysens:85, gallagher:92:pra.46.2012, gallagher:92, kurnaz:95}
(see also Ref.~\cite{beysens:94} and references therein) as well as in
other binary mixtures \cite{narayanan:93, jayalakshmi:97, gruell:97,
guo:188303}. However, only recently, the first \emph{direct}
experimental evidence of the critical Casimir force has been reported
concerning the force between a spherical colloidal particle and a
homogeneous \cite{hertlein:08} or chemically patterned
\cite{soyka:208301, troendle:09} wall.

In the present study we analyze the critical Casimir effect for
non-spherical colloidal particles. In this case, there is not only a
force acting between particles or between a particle and a wall but
there is also a \emph{torque} exerted on the particle. This may lead to
interesting effects such as orientational ordering of non-spherical
colloids in a critical solvent or anchoring of non-spherical particles
at a wall. By varying the temperature the strength of this orientational
interaction can be tuned and by changing chemically the preferences of
the surfaces for the two species forming the solvent one can choose the 
sign of the interaction \cite{sprenger:06, soyka:208301, troendle:09}. 
Motivated by this prospect we therefore extend our previous study of 
critical adsorption at a single non-spherical colloidal particle 
\cite{kondrat:07} to the case that in addition there is a planar wall 
present.

This kind of orientation dependence of fluctuation--induced forces has
recently been studied for quantum mechanical Casimir forces
\cite{rodriguez:08, emig:08}. As far as torque due to critical
fluctuations is concerned, in Ref.~\cite{palagyi:04} the critical
Casimir torque on the confining walls of a wedge has been analyzed.
Based on field-theoretic techniques the interaction of non-spherical
particles, embedded into a solution of long polymers, with a planar wall
has been investigated in the limiting case that the size of the particle
is much smaller than the distance from the wall (``protein limit'') and
which, in turn, is assumed to be much smaller than the correlation
length \cite{eisenriegler:03, eisenriegler:05, eisenriegler:06:204903,
eisenriegler:06:144912}. These latter analyses show, in particular, that 
for a solution of ideal polymer chains the preferred orientation of the
elongated colloidal particle changes from being perpendicular to being
parallel to the substrate surface upon decreasing the particle--wall
distance \cite{eisenriegler:03, eisenriegler:05}, whereas in a solvent
of self-avoiding chains the preferred orientation is the parallel one
for all distances \cite{eisenriegler:06:144912}.

The theoretical interest in the behavior of non-spherical colloidal
particles is matched by an increased experimental interest, even with
application perspectives \cite{harn:07}. Rodlike \cite{liu:03} or
disklike \cite{webe:07} architectures, dumbbell-shaped particles
\cite{john:05,kim:06}, and particles with ellipsoidal shape
\cite{saca:06,hu:08} have been synthesized and characterized. The size
of these particles ranges between 10 nm and 10 $\mu$m. Very recently the
influence of an effective torque exerted by a non-critical one-component
solvent on dumbbell-shaped particles has been revealed using depolarized
light scattering \cite{hoff:08}. For this system, it has been
demonstrated that the addition of small amounts of electrolyte has a
significant impact on the rotational motion and the aggregation
stability of these non-spherical particles, while changes of  the
temperature typically result in only minor changes of the effective
interaction between  colloidal particles because this solvent is not
close to a continuous phase transition. However, as stated above, near a
critical point the effective interaction is expected to exhibit a very
sensitive temperature dependence.

The reminder of the paper is organized as follows. In
Sec.~\ref{sec:model} we define the system under consideration and we
introduce the scaling functions for the critical Casimir interactions.
In order to calculate the force and the torque we use the stress tensor,
as described in Subsec.~\ref{sec:mf:stress_tensor}. The results of
the full mean field calculations are discussed in
Subsecs.~\ref{sec:mf:force} and \ref{sec:mf:torque}. In
Sec.~\ref{sec:comparison} we compare qualitatively the
quantum-electrodynamic Casimir, critical Casimir, and polymer depletion
interactions acting on an ellipsoidal colloid close to a wall. Finally,
in Sec.~\ref{sec:summary} we briefly summarize our results.

\section{Model}
\label{sec:model}

\begin{figure}
\begin{center}
\ifOnecolumn
  \includegraphics*[width=0.5\textwidth]{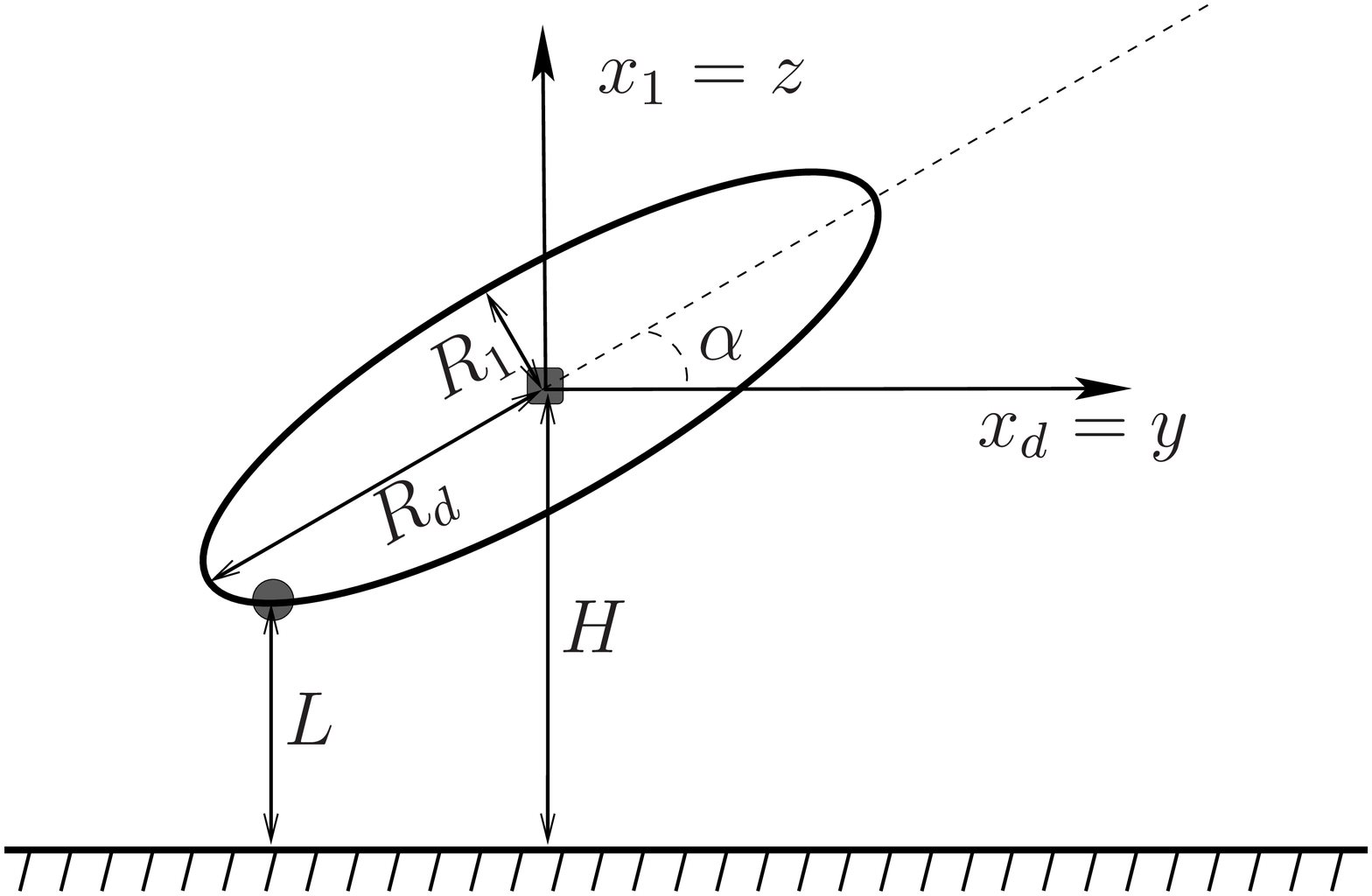}
\else
  \includegraphics*[width=0.5\textwidth]{figs_ellipsoid}
\fi
  \caption{ \label{fig:model}
    Schematic side view of an ellipsoid near a planar wall. The
    ellipsoid semi-axes are $R\equiv R_1 \le R_2 \le \cdots \le R_d$.
    Only the projection onto the $(x_1,\,x_d) \equiv (y,z)$ plane is
    shown. The angle between the long axis of the ellipsoid and the wall
    surface is denoted as $\alpha$, and the closest distance between the
    wall and the surface of the ellipsoid is denoted as $L$, while the
    distance between the center of the ellipsoid and the wall is denoted
    as $H$. The full circle and the square denote two pivots which we
    use to calculate the torque (see Subsec.~\ref{sec:mf:torque}).}
\end{center}
\end{figure}

In order to be able to discuss and to compare the behavior of both 
non-spherical as well as spherical colloidal particles, we describe them
in a unified way by considering a particle with the shape of a
hypercylinder:
\ifOnecolumn
\begin{align}
\label{eq1}
  {\cal K}_d(\{R_i\}) = \left\{{\bf r} = (x_1, x_2, \cdots, x_D)
    \quad \left| \quad \sum\limits_{s=1}^d
	\left(\frac{x_s}{R_s}\right)^2 \le 1\,, \quad d \le D \right.\right\}\,,
\end{align}
\else
\begin{multline}
\label{eq1}
  {\cal K}_d(\{R_i\}) = \Big\{ {\bf r} = (x_1, x_2, \cdots, x_D) \\
    \Big| \quad \sum\limits_{s=1}^d
	\left(\frac{x_s}{R_s}\right)^2 \le 1\,, \quad d \le D \Big\}\,,
\end{multline}
\fi
where $R=R_1 \le R_2 \le \cdots \le R_d$ are the semi-axes of the
hypercylinder (see Fig.~\ref{fig:model}). If $R=R_1 =\cdots = R_{d}$ and
$d=D$, the hypercylinder reduces to a hypersphere. If not all semi-axes
are equal but $d=D$ the hypercylinder is an ellipsoid. It is called a
spheroid if only two semi-axes are different: in $D=3$ one has a prolate
spheroid ($R_1=R_2<R_3$) and an oblate spheroid ($R_1<R_2=R_3$). In the
case $1<d<D$, we have a spheroidal cylinder (or ``spheroido-cylinder''), 
with a $D-d$ dimensional hyperaxis. For reasons of simplicity, and in
order to distinguish the cases of equal and different semi-axes, we
shall use the notion of a disk ($d=2$) and of a $d$-sphere ($d \ge 3$)
if all finite axes are equal (note that they are cylinders for $D>d$),
while the case in which some of the axes are different will simply be
denoted as an ellipse or as a spheroid (or ellipsoid) depending on $d$
(note that they are spheroido-cylinders in $D>d$).

We consider the generalization of $D$ to values different from three
because the upper critical dimension (for the Ising universality class
of the fluids discussed here) is $D^*=4$. We recall that the universal
quantities such as critical exponents and universal scaling functions
calculated within mean field theory (MFT) are exact in $D > D^*$. (For
$D=D^*$ one expects logarithmic corrections which we do not consider
here.) Accordingly, our MFT results (see Sec.~\ref{sec:mf}) can be
interpreted either as exact results in $D > D^*$ or as mean field
approximations for the dimensions $D=3$ or $D=2$.

The effective force $f$ acting on a particle in the direction normal to
the substrate is the negative derivative of the free energy $\mathcal{F}$ of 
the confined fluid with respect to the closest distance $L$ of the particle
from the wall at fixed orientation. Close to a critical point it
decomposes into the sum of an analytical background part and of a
non-analytical (singular) part \cite{krech:91:prl, krech:92:pra:1922,
krech:92:pra:1886} (for further details see
Ref.~\cite{dantchev:07:pre}). According to the scaling behavior
predicted by renormalization group theory the latter (divided by the
``length'' of the $(D-d)$-dimensional hyperaxis of the hypercylinder
along which $\mathcal{K}_d$ is translational invariant), in the case of 
an ellipsoidal particle close to the wall, can be cast into the form 
($s=1,\cdots,d-1$)
\cite{hanke:prl:98}:
\begin{multline}
\label{eq:K}
\ifOnecolumn
  f (L, R_1,\cdots,R_d, \{\alpha_p\}, T) = \frac{k_B T}{L^{D-d+1}}\,
    K_\pm \boldsymbol{\big(} \Theta_\pm=L/\xi_\pm, \Delta=L/R, 
	\{\alpha_p\}; \{\delta_s=R_{s+1}/R\}\boldsymbol{\big)},
\else
  f (L, R_1,\cdots,R_d, \{\alpha_p\}, T) = \frac{k_B T}{L^{D-d+1}}\, \\
    \times K_\pm \boldsymbol{\Big(} \Theta_\pm=L/\xi_\pm, \Delta=L/R,
	\{\alpha_p\}; \\ \{\delta_s=R_{s+1}/R\}\boldsymbol{\Big)},
\fi
\end{multline}
where $\{\alpha_p\}$ are $d (d-1) / 2$ angles determining the
orientation of the hypercylinder, $\xi_\pm=\xi_0^\pm |t|^{-\nu}$ is the
bulk correlation length in the disordered ($+$) and ordered ($-$) phase,
respectively, $t=(T-T_c)/T_c$ with $\nu$ as a standard bulk critical
exponent, and $\xi_0^\pm$ are non-universal amplitudes. $L$ is the
closest distance between the surface of the ellipsoid and the wall (see
Fig.~\ref{fig:model}), $R = R_1$, and $K_\pm$ are dimensionless
universal scaling functions. $f>0$ ($f<0$) corresponds to repulsive
(attractive) forces.

In addition, a non-spherical particle experiences an effective torque,
the components of which are, in general, linear combinations of the
(negative) derivatives of the free energy with respect to the angles
$\{\alpha_p\}$. In the case shown in Fig.~\ref{fig:model}, the torque
has only one  non-zero component describing the rotations within the
$yz$ plane; this component equals the  negative derivative of the free
energy $\mathcal{F}$ with respect to the angle $\alpha$. In general, the
non-analytic contribution to the torque per ``length'' of the
$(D-d)$-dimensional hyperaxis of $\mathcal{K}_d$ takes on the scaling
form
\ifOnecolumn
\begin{align}
\label{eq:T}
  \mathfrak{t}(L, R_1,\cdots,R_d, \{\alpha_p\}, T) = \frac{k_B T}{L^{D-d}}\,
    M_\pm \boldsymbol{\big(}
		\Theta_\pm, \Delta, \{\alpha_p\} ;\{\delta_s\} 
	  \boldsymbol{\big)},
\end{align}
\else
\begin{multline}
\label{eq:T}
  \mathfrak{t}(L, R_1,\cdots,R_d, \{\alpha_p\}, T) = \frac{k_B T}{L^{D-d}} \\
    \times M_\pm \boldsymbol{\big(}
		\Theta_\pm, \Delta, \{\alpha_p\} ;\{\delta_s\} 
	  \boldsymbol{\big)},
\end{multline}
\fi
where $M_\pm$ are the universal scaling functions of the torque. We note
that $\mathfrak{t}$ (and therefore $M_\pm$) is an antisymmetric tensor
of the second rank (the so-called $2$-form). In $D$ dimensions
$\mathfrak{t}$ has, in general, $\tbinom{D}{p=2} =
D!/\boldsymbol{(}2(D-2)!\boldsymbol{)}$ components. However, because of
the translational invariance in $(D-d)$ directions, only $\tbinom{d}{2}
= d(d-1)/2$ of them are non-zero (some of them may as well be zero
depending on additional symmetries of $\mathcal{K}_d$). In $D=3$
dimension the conventional torque pseudo-vector
$\boldsymbol{\mathfrak{t}}$ is obtained by taking the Hodge dual of
$\mathfrak{t}$, i.e., $\boldsymbol{\mathfrak{t}} = * \mathfrak{t}$; in
Euclidian space $\boldsymbol{\mathfrak{t}}_k = (1/2!) \sum_{i,j}
\mathfrak{t}_{ij} \epsilon_{ijk}$, where $\epsilon_{ijk}$ is the
Levi-Civita symbol.

In the literature two common choices are used to define the distance
between the particle and the wall: The surface-to-surface distance $L$
(used, e.g., in Refs.~\cite{hanke:prl:98, schlesener:03, troendle:09})
and the distance $H$ from the center of the particle to the wall (used,
e.g., in Refs.~\cite{eisenriegler:03, eisenriegler:05,
eisenriegler:06:204903, eisenriegler:06:144912, emig:08}), see
Fig.~\ref{fig:model}. Correspondingly, one can also define two different
pivots (points of rotation) with respect to which the particle rotates:
The point on the surface of the particle closest to the wall (denoted by
a circle in Fig.~\ref{fig:model}) in the former case and the center of a
particle (denoted by a square in Fig.~\ref{fig:model}) in the latter
case. While it is obvious that the force does not depend on the
definition of the distance, the torque does depend on the choice of the
pivot. It is easy to see, however, that the values of the torque
corresponding to two different pivots, $\mathfrak{t}^{(1)}$ and
$\mathfrak{t}^{(2)}$, are related by the simple equation (see, c.f.,
also Eq.~(\ref{eq:torque:relation2}))
\begin{align}
\label{eq:torque:relation}
\mathfrak{t}^{(1)} = \mathfrak{t}^{(2)} + 
    \boldsymbol{r}^{(12)} \wedge \boldsymbol{f}, 
\end{align}
where $\boldsymbol{r}^{(12)}=\boldsymbol{r}^{(1)}-\boldsymbol{r}^{(2)}$ 
is the vector connecting the two pivot points located at 
$\boldsymbol{r}^{(1)}$ and $\boldsymbol{r}^{(2)}$, respectively.
$\boldsymbol{f}$ is the vector of the force, and $\wedge$ denotes the
wedge product (a multi-dimensional analogue of the three-dimensional
cross product; its $ij$ component is given by $(\boldsymbol{r}^{(12)}
\wedge \boldsymbol{f})_{ij} = x^{(12)}_i f_j - x^{(12)}_j f_i$, where
$x^{(12)}_i$ is the $i$-th component of $\boldsymbol{r}^{(12)}$. In the
following we shall calculate the scaling functions $K_\pm$ and $M_\pm$
within MFT, and we will also compare and discuss in more detail the
scaling functions for these two choices of the distance and of the pivot
(see Sec.~\ref{sec:mf:force} and
\ref{sec:mf:torque}). 

\section{Mean--field approximation}
\label{sec:mf}

The standard Landau--Ginzburg--Wilson Hamiltonian for critical phenomena
confined to a volume $V$ is given by 
\begin{align} 
\label{eq:hamiltonian}
{\cal H}[\phi] = \int_V dV\,\left(\frac{1}{2}(\nabla \phi)^2+\frac{\tau}{2}\phi^2+
    \frac{u}{24}\phi^4\right)
\end{align}
augmented by boundary conditions \cite{binder:83,diehl:86}. In the case
of a binary liquid mixture near its consolute point the order parameter
$\phi$ is proportional to the difference of the concentrations of its
two species; $\tau$ is proportional to $t$ and $u$ is a coupling
constant which stabilizes ${\cal H}[\phi]$ in the ordered phase. For the
critical adsorption fixed point \cite{diehl:86}, valid for confined
fluids, the boundary conditions are $\phi=+\infty$ (or $\phi=-\infty$)
at the surface of the colloidal particle and at the wall, to which we
refer to in the following as the ``$+$'' (or ``$-$'') boundary
condition. The semi-axes \{$R_i$\} of the hypercylinder (see
Fig.~\ref{fig:model}) introduce additional length scales which might
come into play via coupling constants of additional surface terms in the
effective Hamiltonian \cite{diehl:86}. However, on the basis of power
counting one concludes that such terms are irrelevant at the ordinary
transition (corresponding to the Dirichlet boundary conditions), where
the surface enhancement coupling of the term proportional to $\phi^2$ at
the surface asymptotically dominates $1/R_i$ contributions to couplings
of symmetry-preserving  boundary terms. On the same footing we expect
that the distinctive feature  of the normal transition, i.e., the
occurrence of symmetry breaking with a resulting asymptotic divergence
of the order parameter at the surface is asymptotically not affected by
curvatures; curvatures might perhaps influence the cross-over between
the ordinary and the normal transitions. Accordingly, we do not expect
the aforementioned asymptotic boundary  conditions to be modified by
curvature.

Within MFT, the fluctuations  of the order parameter $\phi$
are neglected and only the order parameter configuration with the
largest statistical weight, $m=\sqrt{u/6}\langle \phi \rangle$, is taken
into account. Minimization of Eq.~(\ref{eq:hamiltonian}) leads to the 
Euler--Lagrange equation
\begin{align} 
\label{eq:EL}
\Delta m=\tau m +m^3\,.
\end{align}
Equation (\ref{eq:EL}) is solved numerically as function of $\tau$ using
the finite element method (see, e.g., Ref.~\cite{fem:book}). We consider
only hypercylinders ${\cal K}_d(\{R_i\})$ with $d=2$ and $d=3$, for
which the problem is effectively two- and three-dimensional,
respectively (for ${\cal K}_{d\ge 4}$ one has a four and higher
dimensional problem which is difficult to solve numerically). For
reasons of simplicity we also restrict our considerations to the case of
only two different semi-axes, $R = R_1 =\cdots =R_{d-1} < R_d$, i.e., we
consider a prolate spheroid in spatial dimension $d$ (which is a
spheroido-cylinder in spatial dimension $D>d$). We expect that the
results for an oblate spheroid do not differ qualitatively. In
Eq.~(\ref{eq:EL}) $\tau$ can be expressed in terms of the bulk
correlation length $\xi_\pm$, which governs the exponential decay of the
two-point correlation function in the bulk: $\tau=\xi_+^{-2}$ for $\tau
> 0$ and $\tau = - \xi_-^{-2} / 2$ for $\tau < 0$.

We note that the contribution from the square gradient term  in
Eq.~(\ref{eq:hamiltonian}) does not vanish in the case of the 
ellipsoid-wall geometry with symmetry-breaking boundary conditions.
Therefore, one does not expect an additional thermodynamic length 
\cite{zinn:89} to emerge and to affect finite-size scaling in 
dimensions $D > D^*$. Such a thermodynamic length naturally  emerges in
the finite-size scaling analysis for $D > D^*$ if the  standard
mean-field theory in a finite volume exhibits an isolated zero mode
which becomes ``massless''  at the bulk critical point due to the
absence of a nonvanishing contribution from the square  gradient term in
Eq.~(\ref{eq:hamiltonian}).

\subsection{The stress tensor}
\label{sec:mf:stress_tensor}

By using the stress tensor, the force and the torque can be calculated
directly from the order parameter profile  $m({\mathbf r})$. This has
the advantage that one avoids the numerical difficulties of calculating
differences of free energies which attain large values due to the
divergence of the order parameter profile at the surfaces. 

We consider an infinitesimal, local coordinate transformation
($k=1,\cdots,D$):
\begin{align}
\label{eq:coord}
x_k' = x_k + \sum_j X_{kj}\, \delta \omega_j({\mathbf r}),
\end{align}
where the meaning of the index $j$ depends on the kind of transformation
and is specified below. The linear response of a system to such a
coordinate transformation is
\begin{align}
\delta{\cal H} = \int_V d V\,\sum_{j,k} \theta_{j k}({\mathbf r})\,
		\frac{\partial\, \delta \omega_j}{\partial x_k},
\end{align}
where ($\partial_n m= \partial m / \partial x_n$)
\begin{align}
\label{eq:Theta}
\theta_{j k}({\mathbf r}) = \frac{\partial\,{\cal L}}{\partial\, (\partial_k m)} 
    \sum_{n} (\partial_n m) X_{jn}
    - {\cal L} X_{jk}\,.
\end{align}
In Eq.~(\ref{eq:Theta}) ${\cal L}(m,\partial m)$ is the integrand of the
Hamiltonian ${\cal H}$ given by Eq.~(\ref{eq:hamiltonian}). In order to
calculate the force, we use the coordinate transformation with
$X_{jk}=\delta_{jk}$ ($j,k=1,\cdots,D$) and $\delta\omega_z({\mathbf r})
= a$ for ${{\mathbf r}}\in V_0$ and $\delta\omega_k({{\mathbf r}}) = 0$
otherwise, where $V_0$ is a generalized hypercylinder enclosing
$\mathcal{K}_{d}$. Accordingly, the $z$-component of the force, which is
the only non-zero component (see Fig.~\ref{fig:model}), divided by the 
``length'' $\ell = \int d^{D-d}x$ of the ($D-d$)-dimensional hyperaxis
of $\mathcal{K}_d$ is given by \cite{schlesener:03}
\begin{align}
\label{eq:force}
\frac{f}{k_B T} = - \frac{1}{\ell} \frac{\partial\, \delta{\cal H}}{\partial a} 
    = \frac{1}{\ell} \oint_{S} d S\, \sum_{k=1}^{D} {\cal T}_{zk}\, \hat{n}_k 
    = -\frac{1}{\ell} \frac{\partial\, \mathcal{F}}{\partial L} ,
\end{align}
where within MFT $\mathcal{F} = \min_\phi \left( \mathcal{H}[\phi]
\right)$. $S$ denotes the surface of $V_0$, $\hat{n}_k$ is the $k$-th
component of its unit outward normal, $dS=d^{D-1}x$, and ${\cal T}_{jk}$
is the conventional stress tensor:
\begin{align}
  {\cal T}_{jk} = \frac{\delta {\cal L}}{\delta (\partial_k m)} \partial_j m 
  -  \delta_{jk} {\cal L}.
\end{align}
One often   adds to ${\cal T}$ the so-called ``improvement'' term (see,
e.g., Refs.~\cite{schlesener:03, sprenger:06, maciolek:06,
maciolek:07}),
\begin{align}
  \mathcal{I}_{kl} = \frac{1}{4}\frac{D-2}{D-1}\, 
    \big[ \partial_k\partial_l - \delta_{kl}\Delta \big] m^2,
\end{align}
which ensures the scale and conformal invariance of the stress tensor at
the critical point and renders it renormalizable \cite{callan:70,
collins:76:prl, collins:76, brown:80}. However, the contributions from 
this term to `observable quantities' like the force or the torque
vanish. For instance, in the case of the force one  obtains with the
help of the Gauss--Ostrogradsky theorem (using here the summation
convention) $\oint_S \mathcal{I}_{lk} \hat n_l dS = \int_{V_0}
\partial_l \mathcal{I}_{lk} dV \equiv 0$ because $\partial_l
\mathcal{I}_{lk} = 0$.

In order to calculate the torque, the coordinate transformation
(\ref{eq:coord}) is chosen to take the form of an infinitesimal
rotation, i.e.,
\begin{align}
\label{eq:coord:rot}
x_k' = x_k + \sum_{\substack{n=1\\ n\ne k}}^{D} x_n\,\delta\omega_{nk} 
    = x_k + \sum_{\substack{l,n=1\\ n<l}}^{D} X_{k,nl}\, \delta \omega_{nl},
\end{align} 
where $X_{k,nl} = x_l\delta_{kn} - x_n\delta_{kl}$ and the antisymmetry
of $\delta\omega_{nk}$ has been used. Note that here the index $j$ in
Eq.~(\ref{eq:coord}) consists of two indices, $j \to (ln)$, which denote
the rotation plane. Now we choose $\delta\omega_{yz}({\mathbf r}) =
\alpha$ for ${\mathbf r} \in V_0$ and $\delta\omega_{nk}({\mathbf r})=0$
otherwise (see Fig.~\ref{fig:model}). This renders the (only non-zero)
component of the torque (per ``length'' of the ($D-d$)-dimensional
hyperaxis of $\mathcal{K}_d$):
\begin{align}
\label{eq:torque}
\frac{\mathfrak{t}_{yz}}{k_BT} = - \frac{1}{\ell} \frac{\partial\, \delta{\cal H}}{\partial \alpha}  
    = \frac{1}{\ell} \oint_{S} d S\, \sum_{k=1}^{D} {\cal M}_{k,yz}\,\hat{n}_k
    = -\frac{1}{\ell} \frac{\partial\, \mathcal{F}}{\partial \alpha},
\end{align}
where the ``angular momentum'' tensor ${\cal M}_{k,nl}$ is
\begin{align}
\label{eq:momentum}
{\cal M}_{k,nl} = \left(x_l - x_l^{(1)}\right) {\cal T}_{kn} 
		- \left(x_n - x_n^{(1)}\right) {\cal T}_{kl},
\end{align}
with  $x_l^{(1)}$ being the $l$ component of the position vector of the
pivot with respect to which the ellipsoid rotates. If we choose
$x_l^{(1)} = x_l^{(2)} + x_l^{(12)}$, where $x_l^{(2)}$ is the $l$
component of the position vector of another pivot and $x_l^{(12)}$ is
the $l$ component of the vector $ \boldsymbol{r}^{(12)} =
\boldsymbol{r}^{(1)}-\boldsymbol{r}^{(2)}$ connecting the two pivots,
then by using Eqs.~(\ref{eq:force}) and (\ref{eq:torque}) we obtain
\begin{align}
\label{eq:torque:relation2}
\mathfrak{t}_{nl}^{(1)} = \mathfrak{t}_{nl}^{(2)} 
	+ x^{(12)}_n f_l - x^{(12)}_l f_n,
\end{align}
where $f_l$ is the $l$ component of the force $\boldsymbol{f}$ (we
recall that in our case only $f_z \equiv f \ne 0$), and
$\mathfrak{t}^{(i)}_{nl}$ denotes the $(nl)$ component of the torque
corresponding to the $i$-th pivot. Equation (\ref{eq:torque:relation2})
is the component version of Eq.~(\ref{eq:torque:relation}).

\subsection{Casimir force}
\label{sec:mf:force}

\begin{figure}[t]
 \begin{center}
\ifOnecolumn
   \includegraphics*[width=0.45\textwidth]{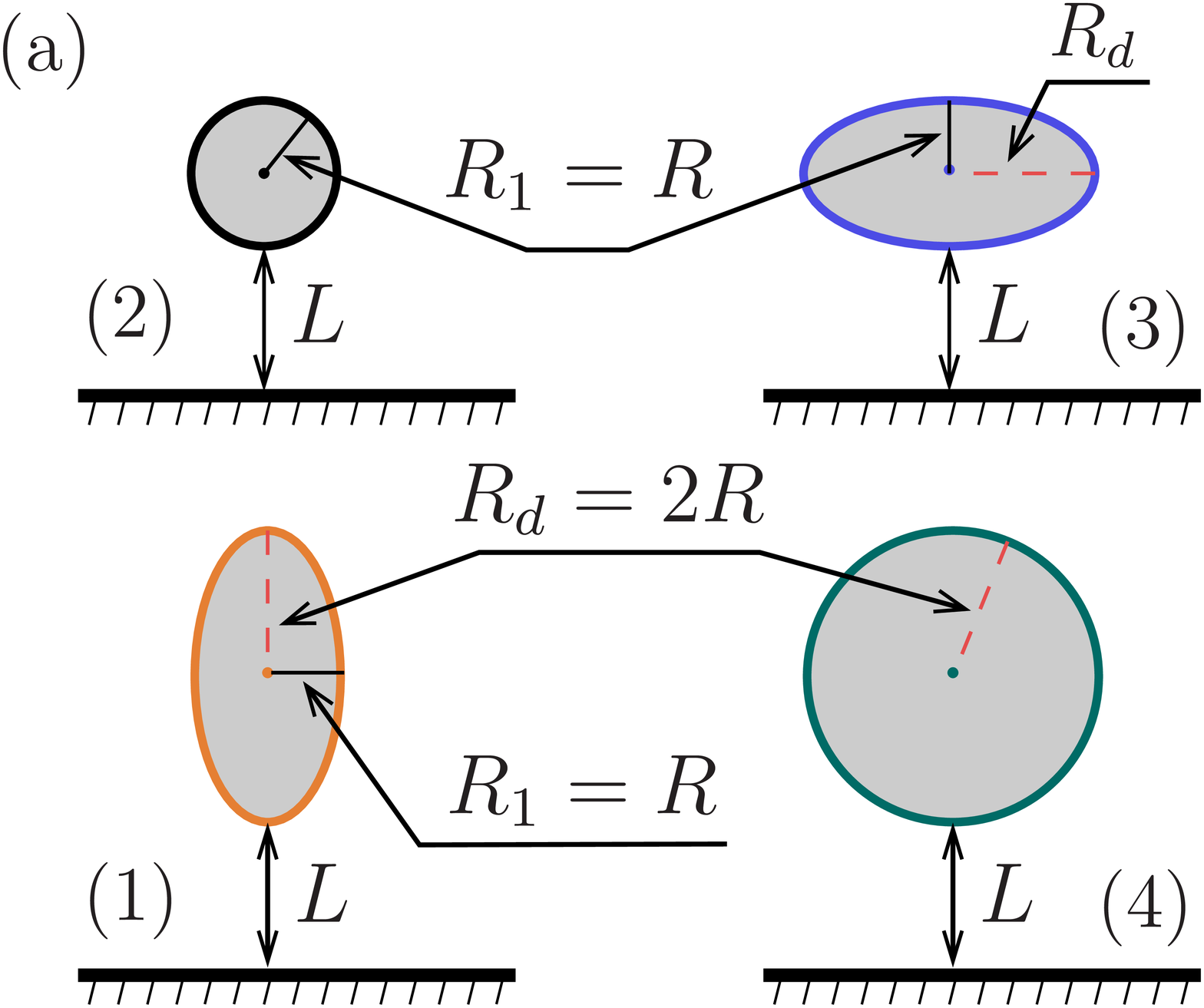}
   \includegraphics*[width=0.47\textwidth]{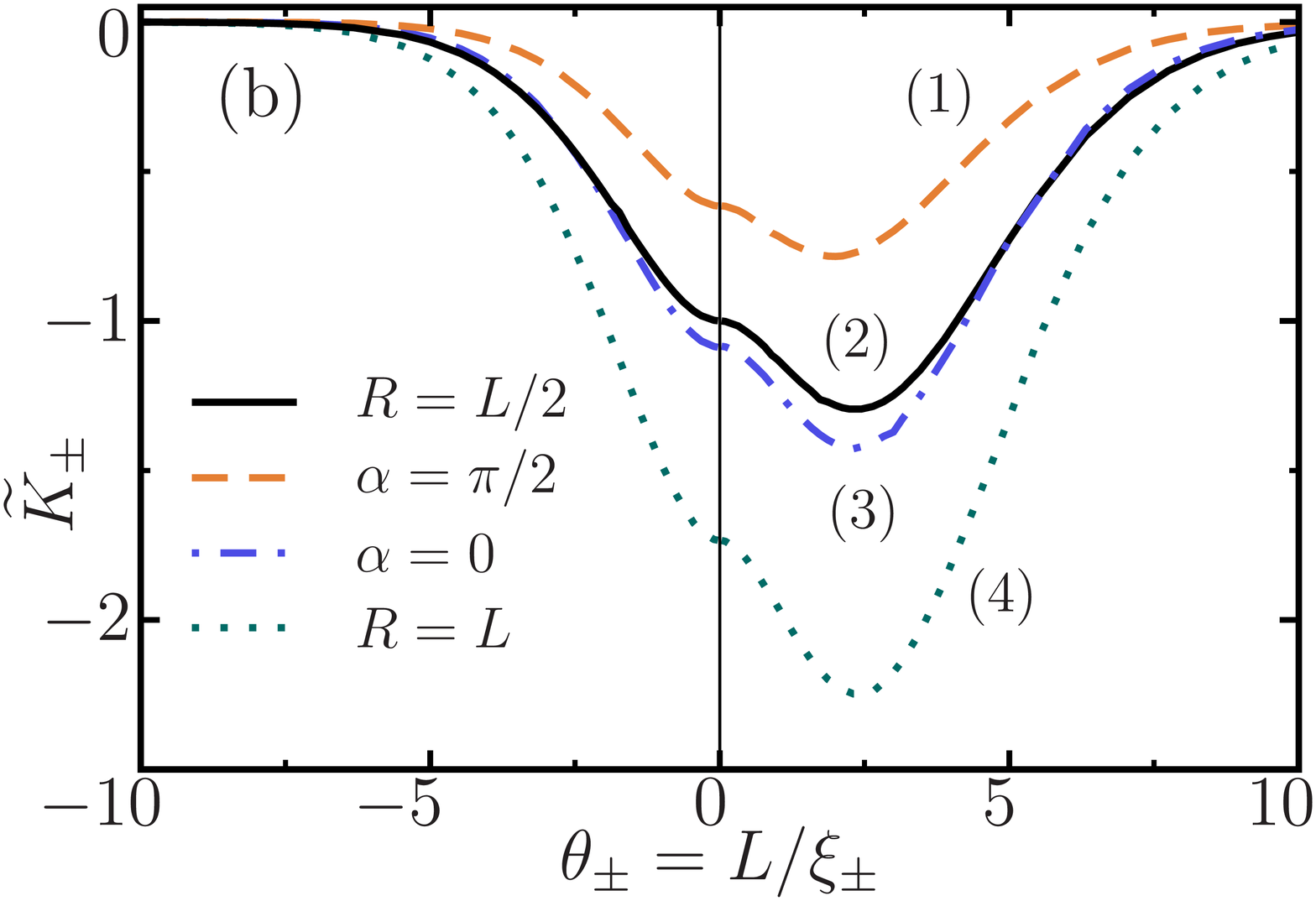}
\else
   \includegraphics*[width=0.4\textwidth]{figs_comparison}
   \includegraphics*[width=0.4\textwidth]{agrs_comparison}
\fi
   \caption{(color online) (a) Two spheres ((2) and (4)) and one
   spheroid with perpendicular ($\alpha=\pi/2$; (1)) and parallel
   ($\alpha=0$; (3)) orientation at the same surface-to-surface distance
   $L$ from the wall. (b) The universal scaling functions $\widetilde
   K_\pm= K_\pm(\Theta_\pm, \Delta=2, \alpha;\,\delta=2) / | K_+
   (\Theta_+=0, \Delta=2;\, \delta=1)|$ as a function of the rescaled
   reduced temperature  $\Theta_\pm = L/\xi_\pm = |t|^\nu L /
   \xi_0^\pm$, where $\delta=R_d/R$. The curves (2) and (4) correspond
   to ($d=3$)-spheres with radii   $R=R_1=L/2$ and $R=R_d=2R_1=L$,
   respectively. The curves (1) and (3)  correspond to a \emph{prolate}
   spheroid with $R=R_1=R_2<R_{d=3}$  (spheroido-cylinder ${\cal K}_3$
   in spatial dimension $D \ge 4$) for  $\Delta=L/R=2$,
   $\delta=R_{d=3}/R=2$, and the orientations $\alpha=\pi/2$ and 
   $\alpha = 0$, respectively, where $\alpha$ is the angle between the
   main  axis of the spheroid and the wall (see Fig.~\ref{fig:model}).
   In all cases $++$ boundary conditions are imposed. The force is
   expressed in units of the absolute value of the Casimir force for the
   sphere with radius $R=L/2$ at the critical point ($\Theta_+=0$) and
   for $++$ boundary conditions (accordingly, at $\Theta_+ =0 $ line (2)
   is normalized to $-1$).
   \label{fig:force:distance_comp}}
 \end{center}
\end{figure}

\begin{figure}[ht]
 \begin{center}
\ifOnecolumn
   \includegraphics*[width=0.47\textwidth]{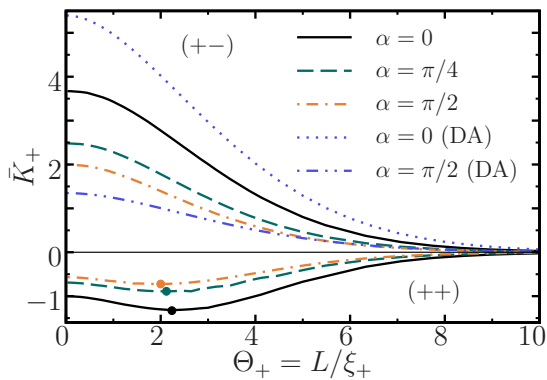}
\else
   \includegraphics*[width=0.4\textwidth]{agrs_force}
\fi
   \caption{(color online) The universal scaling function $\bar K_+= K_+
   (\Theta_+, \Delta=2, \alpha;\, \delta=2) / |K_+ (\Theta_+=0,
   \Delta=2, \alpha=0;\, \delta=2)|$ as a function of the rescaled
   reduced temperature $\Theta_+ = L/\xi_+ = t^\nu L/\xi_0^+$ for a
   prolate spheroid $R=R_1=R_2<R_3$ (spheroido-cylinder
   $\mathcal{K}_{d=3}$ in $D>d$) for $\Delta=L/R = 2$ and $\delta =
   R_3/R=2$, and for three values of the angle $\alpha$ between the main
   axis of the spheroid and the wall (see Fig.~\ref{fig:model}). The
   positive (negative) curves correspond to $+-$ ($++$) boundary
   conditions on the spheroid and the wall leading to repulsion
   (attraction). The force is expressed in units of the absolute value
   of the Casimir force for $\alpha = 0$ at the critical point
   ($\Theta_+=0$) and for  $++$ boundary conditions. Accordingly, at
   $\Theta_+=0$ the bottom curve is normalized to $-1$. (Note that here
   the normalization differs from the one used in
   Fig.~\ref{fig:force:distance_comp}.) The dots denote the minima of
   the scaling function. Dotted and dash--double dotted lines show the
   Derjaguin approximation (DA) for $\alpha=0$ and $\alpha=\pi/2$,
   respectively, and for $+-$ boundary conditions. According to
   Eqs.~(\ref{eq:derjaguin}) and (\ref{eq:curvature}) these two curves
   differ only by an overall scale factor $\delta^2 = 4$.
   \label{fig:force:distance}}
 \end{center}
\end{figure}

We first study the influence of the anisotropy of the particle on the
Casimir force for $D=4$. In Fig.~\ref{fig:force:distance_comp}(b) we
compare the universal scaling functions $K_\pm$, as functions of the
rescaled distance $\Theta_\pm = L/\xi_\pm$, for a spheroidal particle
${\cal K}_{d=3}$ and for two ($d=3$)-spheres (see Eq.~(\ref{eq1})) in
the case of $++$ boundary conditions which lead to attractive forces. In
particular, we compare the two extreme orientations $\alpha = 0$ and
$\alpha = \pi/2$ of a spheroid (i.e., the particle is oriented parallel
and perpendicular to the wall, respectively, see Fig.~\ref{fig:model})
with two ($d=3$)-spheres: a small one with its radius equal to the
smallest semi-axis ($R=R_1$), and a bigger one with its radius equal to
the major semi-axis of the spheroid ($R=R_d=2R_1$). It is interesting
that the force is stronger for the small sphere ((2) in
Fig.~\ref{fig:force:distance_comp}(a)) than for the spheroid oriented
perpendicular to the wall ((1) in
Fig.~\ref{fig:force:distance_comp}(a)). Indeed, within the Derjaguin
approximation \cite{derjaguin:34, derjaguin:75} the force is inversely
proportional to the square root of the Gaussian curvature $G =
\kappa^{(1)} \kappa^{(2)}$ of the particle surface at the point closest
to the wall (c.f., Eq.~(\ref{eq:derjaguin})); $\kappa^{(i)}$ are the
principle curvatures.  The square root of the Gaussian curvature of the
sphere is $G_{\mathrm{sphere}}^{1/2} = 1/R_1$ while the principle
curvatures of the spheroid at its elongated edge ($\alpha = \pi/2$) are
$\kappa^{(1)}_{\mathrm{spheroid}} =
\kappa^{(2)}_{\mathrm{spheroid}} = R_d/R_1^2$ so that
$G_{\mathrm{spheroid}}^{1/2}(\alpha=\pi/2) = R_d/R_1^2$, which is larger
than $G_{\mathrm{sphere}}^{1/2}$ if $R_d > R_1$ (which is the present
case, see Fig.~\ref{fig:force:distance_comp}(a)). For $\alpha = 0$ ((3)
in Fig.~\ref{fig:force:distance_comp}(a)) one has
$G_{\mathrm{spheroid}}^{1/2}(\alpha=\pi/2) = 1/R_d$ which equals
$G_{\mathrm{sphere}}^{1/2}= 1/R_d$ for (4) in
Fig.~\ref{fig:force:distance_comp}(a). Thus the configurations (3) and
(4) in Fig.~\ref{fig:force:distance_comp}(a) have the same Gaussian
curvature at the point closest to the wall so that within the Derjaguin
approximation given by, c.f., Eq.~(\ref{eq:derjaguin}) both cases lead
to the same critical Casimir force. Therefore the difference between the
curves (3) and (4) in Fig.~\ref{fig:force:distance_comp}(b)  highlights
the shortcomings of the Derjaguin approximation
(Eq.~(\ref{eq:derjaguin})) for this geometrical set-up. The scaling
functions  for intermediate orientations and for $++$ as well as $+-$
boundary conditions  are shown in Fig.~\ref{fig:force:distance}. The
minima of the scaling functions for $++$ boundary conditions occur at
$T>T_c$. They move closer to $T_c$ upon changing the orientation of the
particle from  parallel to perpendicular.

It is instructive to compare our full MFT results with the Derjaguin (or
proximity force) approximation \cite{derjaguin:34, derjaguin:75}, which
is applicable for large particles close to the wall. In lowest order in
curvatures one obtains
\ifOnecolumn
\begin{align}
\label{eq:derjaguin}
  K_\pm(\Theta, \Delta, \alpha;\, \delta) = 
    \frac{(2\pi)^{(d-1)/2}}{\Gamma\boldsymbol{(}(d+1)/2\boldsymbol{)}}\,
	\widetilde{G}^{-1/2}(\Delta,\delta, \alpha)
	\int_1^{\infty} \frac{K^{(\parallel)}_{\pm}(\Theta u)}{u^D}(u-1)^{(d-3)/2} du,
\end{align}
\else
\begin{multline}
\label{eq:derjaguin}
  K_\pm(\Theta, \Delta, \alpha;\, \delta) = 
    \frac{(2\pi)^{(d-1)/2}}{\Gamma\boldsymbol{(}(d+1)/2\boldsymbol{)}}\,
	\widetilde{G}^{-1/2}(\Delta,\delta, \alpha) \\
	\times \int_1^{\infty} \frac{K^{(\parallel)}_{\pm}(\Theta u)}{u^D}(u-1)^{(d-3)/2} du,
\end{multline}
\fi
where $K^{(\parallel)}_{\pm}$ is the scaling function for the
plate-plate geometry, $\Gamma(x)$ is the Gamma function, and $\widetilde
G(\Delta,\delta, \alpha)=L^{d-1} \prod_{s=1}^{d-1} \kappa^{(s)}$ (where
$\kappa^{(s)}$ are the principle curvatures) is the dimensionless
Gausian curvature of the particle surface at the point closest to the
wall. In the case of the spheroid $\mathcal{K}_{d=3}$ one has
\begin{align}
\label{eq:curvature}
  \widetilde G \boldsymbol{(}\Delta,\delta, \varphi(\alpha)\boldsymbol{)} = 
    \frac{\Delta^{2}\delta^2}
	    {\boldsymbol{(}1+(\delta^2-1)\cos^2\varphi(\alpha)\boldsymbol{)}^2}\,,
\end{align}
where $\varphi$ is the parametric latitude denoting the position on the
spheroid surface parameterized as $(R_1 \cos\varphi\cos\psi,\,
R_1\cos\varphi\sin\psi, \,R_d\sin\varphi)$ with $\psi$ as the longitude.
The value of $\varphi$ at the point closest to the wall depends on
$\alpha$; $\varphi(\alpha=0)=0$ and $\varphi(\alpha=\pi/2)=\pi/2$.

\begin{figure}[t]
  \begin{center}
\ifOnecolumn
   \includegraphics*[width=0.47\textwidth]{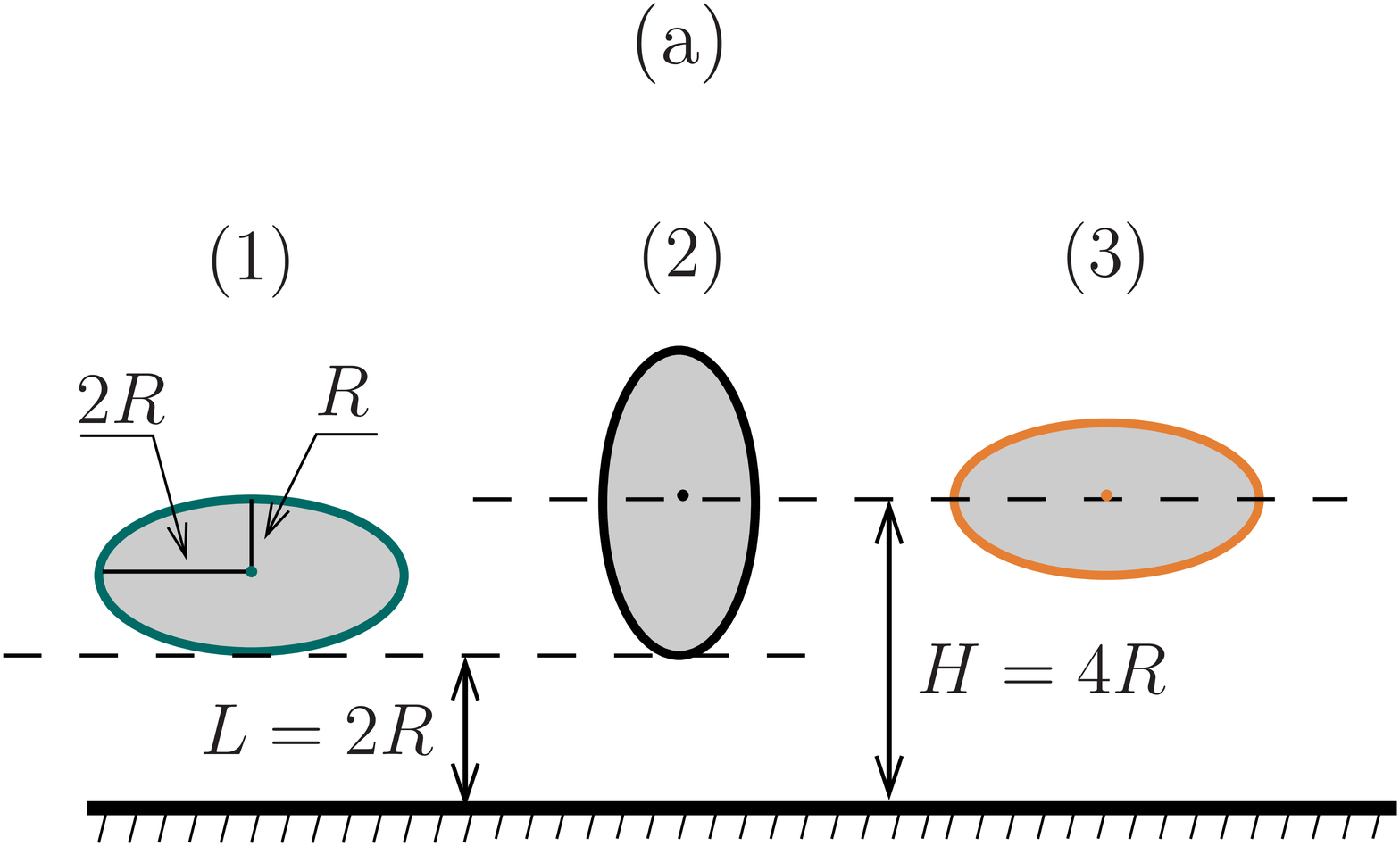}
   \includegraphics*[width=0.47\textwidth]{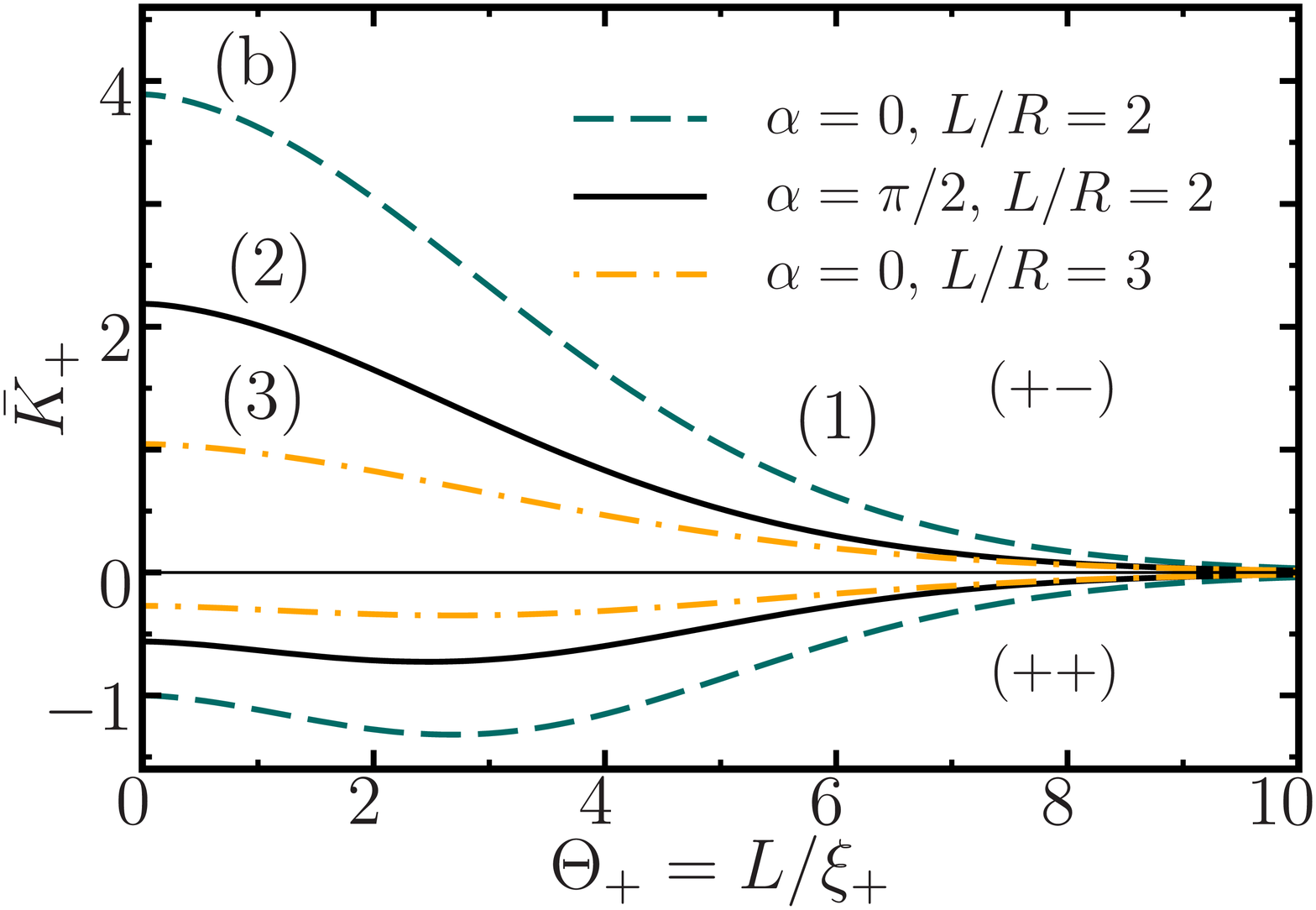}
\else
   \includegraphics*[width=0.4\textwidth]{figs_comparison_HL}
   \includegraphics*[width=0.4\textwidth]{agrs_comparison2}
\fi
    \caption{(color online) (a) Comparison between three ellipse
    configurations such that the configurations (1) and (2) have the
    same closest distance $L=2R$, while configurations (2) and (3) have
    the same distance $H=4R$ between the center of the particle and the
    wall. For the configuration (1) $H=3R$ and for the configuration (3)
    $L=3R$. (b) The universal scaling function $\bar K_+= K_+(\Theta_+,
    \Delta, \alpha;\, \delta=2) / |K_+(\Theta_+=0, \Delta=2, \alpha=0;\,
    \delta=2)|$ as a function of the reduced temperature
    $\Theta_+=L/\xi_+$ for an ellipse ($d=2$). The dashed lines (1)
    represent the results for an ellipse oriented parallel to the wall
    ($\alpha=0$, $L/R=2$, $H/R=3$) at \emph{the same closest distance}
    between its surface and the wall as an ellipse which is oriented
    perpendicular to the wall ($\alpha=\pi/2, L/R=2, H/R=4$, solid lines
    (2)). The dash-dotted lines (3) represent the results for an ellipse
    oriented parallel to the wall ($\alpha=0$, $L/R=3, H/R=4$) and at
    the same distance between its \emph{center of mass} and the wall as
    compared to the ellipse which is oriented perpendicular to the wall
    (solid lines (2)). The positive (negative) curves correspond to $+-$
    ($++$)  boundary conditions on the ellipse and at the wall. The
    force is expressed in units of  the absolute value of the Casimir
    force for $\alpha=0$ and $\Delta=L/R=2$ at the critical point
    ($\Theta_+=0$) and for $++$ boundary conditions.}
    \label{fig:comparison2}
\end{center} 
\end{figure}

The results within the Derjaguin approximation (using the mean field
scaling function $K^{(\parallel)}_{\pm}$)  are shown in
Fig.~\ref{fig:force:distance} by dotted and dash--double dotted lines.
In order to avoid overloading Fig.~\ref{fig:force:distance}, the
Derjaguin results are presented only for two angles, $\alpha = 0$ and
$\alpha=\pi/2$. For these two angles the Derjaguin results differ only
by an overall scale factor $\delta^2$ with $\delta^2=4$ here. For both
angles the discrepancy with the full mean field results is rather large.
The reason is that in the case studied here the ratio $\Delta=L/R=2$,
while the Derjaguin approximation is supposed to be valid in the limit
of small distances (or big particles), i.e., for $\Delta\to 0$. It is
interesting to note that the Derjaguin approximation underestimates the
strength of the force for $\alpha=\pi/2$ and overestimates it for
$\alpha=0$.

For the comparison of the Casimir force acting on ellipsoids with
various orientations and at a fixed distance from the wall in
$z$-direction, there are two interesting choices for this distance: The
closest distance $L$ between the wall and the surface  of the ellipsoid,
and the distance $H$ between the center of the ellipsoid and the  wall
(see Fig.~\ref{fig:model}). Figure \ref{fig:comparison2}(b) displays the
corresponding scaling function for an ellipse  ($d=2$) which is oriented
perpendicular to the wall ($\alpha=\pi/2, L/R=2$, $H/R = 4$, solid 
lines; (2) in Fig.~\ref{fig:comparison2}) together with the scaling
function for an ellipse which is oriented parallel  to the wall at the
same distance $L$ ($\alpha=0$, $L/R=2$, $H/R=3$, dashed lines; (1) in
Fig.~\ref{fig:comparison2}) and the scaling function for an ellipse
which is oriented parallel to the wall at the  same distance $H$ 
($\alpha=0$, $L/R=3$, $H/R=4$, dash-dotted lines; (3) in
Fig.~\ref{fig:comparison2}). If the ellipse is orientated  perpendicular
to the wall (solid lines), the magnitude of the Casimir force is smaller
than if the ellipse is parallel to the wall at the same closest distance
between  the wall and the surface of the ellipse (dashed lines).
Compared to an ellipse  which is parallel to the wall at the same
distance between the center of the ellipse and the wall (dash-dotted
lines) the magnitude of the Casimir force acting on the perpendicular
oriented ellipse is larger (solid lines). We have confirmed numerically
that these relations  between the strengths of the fluctuation induced
forces are valid for arbitrary values  of $\Delta=L/R$ including the
limits $\Delta \to 0$ and $\Delta\gg 1$. In the  latter limit these
results are in agreement with the predictions of a small particle 
operator expansion \cite{eisenriegler:04} which can be used near the
critical point ($\Theta_+\to 0$) as long as all semi-axes of the ellipse
are much smaller than all other lengths such  as the correlation length
or the distance between the surface of the particle  and the wall.

\begin{figure}[ht]
 \begin{center}
\ifOnecolumn
   \includegraphics*[width=0.47\textwidth]{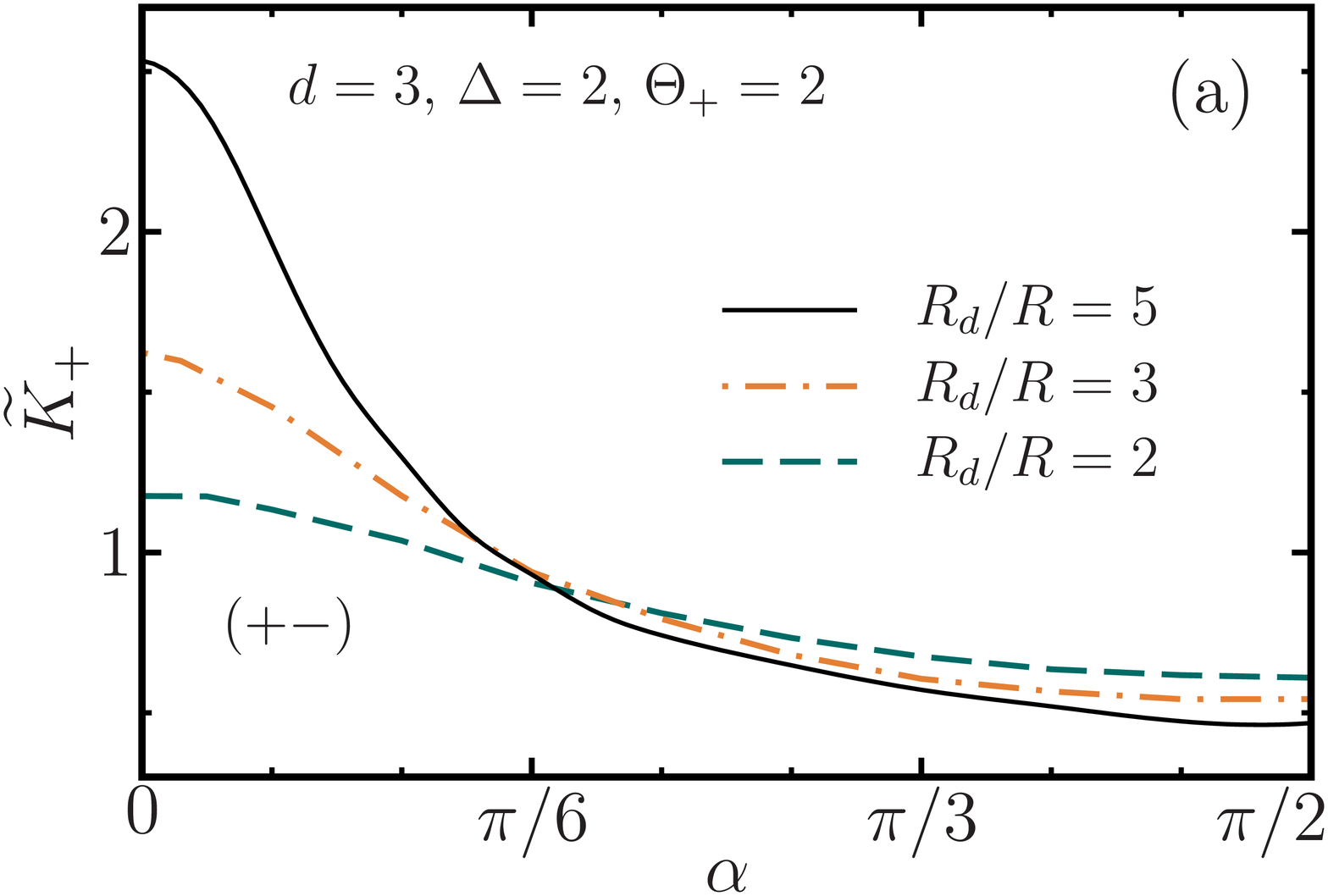}
   \includegraphics*[width=0.47\textwidth]{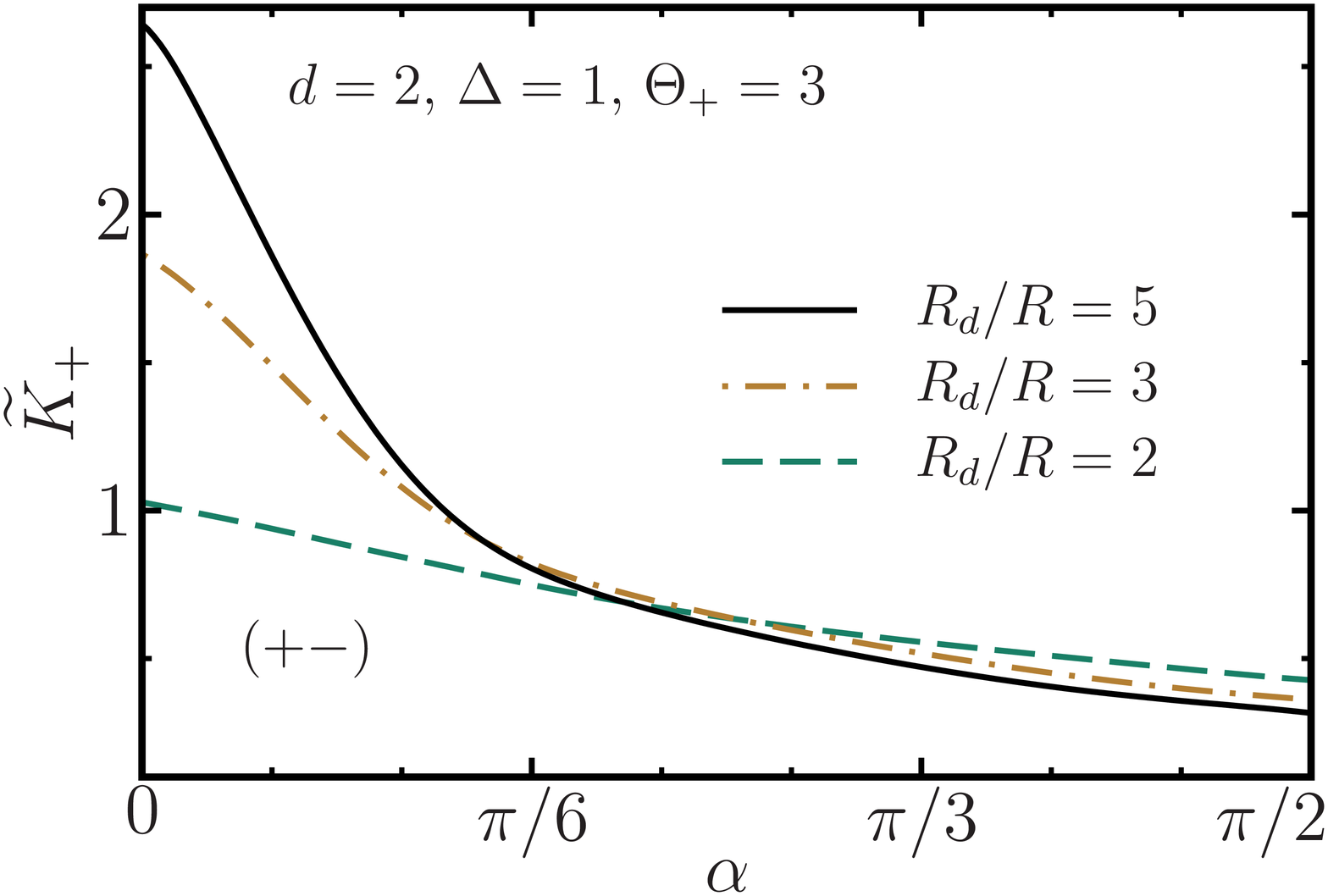}
\else
   \includegraphics*[width=0.4\textwidth]{agrs_d=3-force}
   \includegraphics*[width=0.4\textwidth]{agrs_d=2-force}
\fi
   \caption{(color online) The universal scaling function $\widetilde
   K_+ = K_+(\Theta_+, \Delta, \alpha;\, \delta) / K_+( \Theta_+=0,
   \Delta;\, \delta=1)$ for $+-$ boundary conditions as a function of
   the angle $\alpha$ between the main axis of the spheroid and the wall
   (see Fig.~\ref{fig:model}) for three values of the ratio
   $\delta=R_d/R$. In (a) $d=3$,  $\Theta_+ = L/\xi_+ = 2$, and $\Delta
   = L/R = 2$, while in (b)  $d=2$,  $\Theta_+ = 3$, and $\Delta = 1$.
   The force is expressed in terms of the Casimir force at the critical
   point for $+-$ boundary conditions for a sphere ($d=3$) with radius
   $R=R_1= L/2$ in (a) and for a disk ($d=2$) with radius $R=L$ in (b).
   All curves correspond to keeping temperature and the \emph{minimal
   distance} $L$ fixed upon varying the orientation (see
   Fig.~\ref{fig:model}).
    \label{fig:force:angle}}
 \end{center}
\end{figure}

In Fig.~\ref{fig:force:angle} the scaling function $K_+$ with a suitable
normalization is plotted as a function of the angle $\alpha$ between the
main axis of the spheroid and the wall (see Fig.~\ref{fig:model}) for a
fixed rescaled minimal distance $L/\xi_+ = \mathrm{const}$. As expected,
the force is larger for more elongated colloids ($R_d/R_1$ large) but
only for small $\alpha$ (i.e., if the colloids are almost parallel to
the wall). If the colloids are tilted more towards the perpendicular
orientation ($\alpha \gtrsim 30^\circ$), the opposite trend is observed.
As discussed at the beginning of this subsection, this is due to the
shapes of the prolate spheroids: for the perpendicular orientation
$G^{-1/2}_{\mathrm{spheroid}} (\alpha=\pi/2) = R^2/R_d$ for $d=3$ and
$G^{-1/2}_{\mathrm{ellipse}} (\pi/2) = R/R_d^{1/2}$ for $d=2$, which
decrease upon increasing $R_d$, whereas for the parallel orientation one
has $G^{-1/2}_{\mathrm{spheroid}} (0) = R_d$ for $d=3$ and
$G^{-1/2}_{\mathrm{ellipse}} (0) = R_d / R^{1/2}$ for $d=2$ which
increase upon increasing $R_d$. The scaling functions for the
hypercylinders ${\cal K}_3$ (Fig.~\ref{fig:force:angle}(a)) and ${\cal
K}_2$ (Fig.~\ref{fig:force:angle}(b)) are rather similar. 

\subsection{Casimir torque}
\label{sec:mf:torque}

\begin{figure}[!ht]
 \begin{center}
\ifOnecolumn
   \includegraphics*[width=0.47\textwidth]{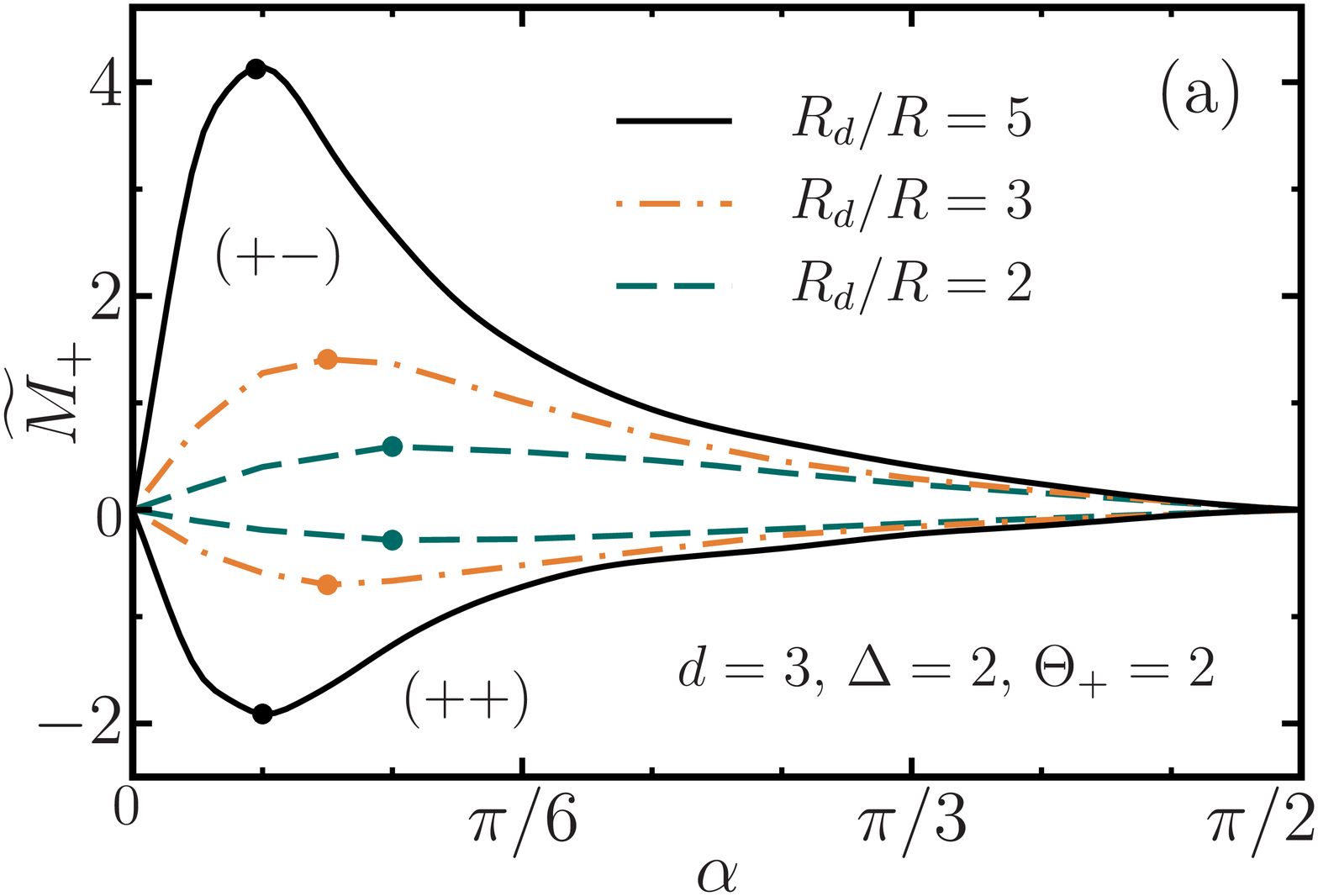}
   \includegraphics*[width=0.47\textwidth]{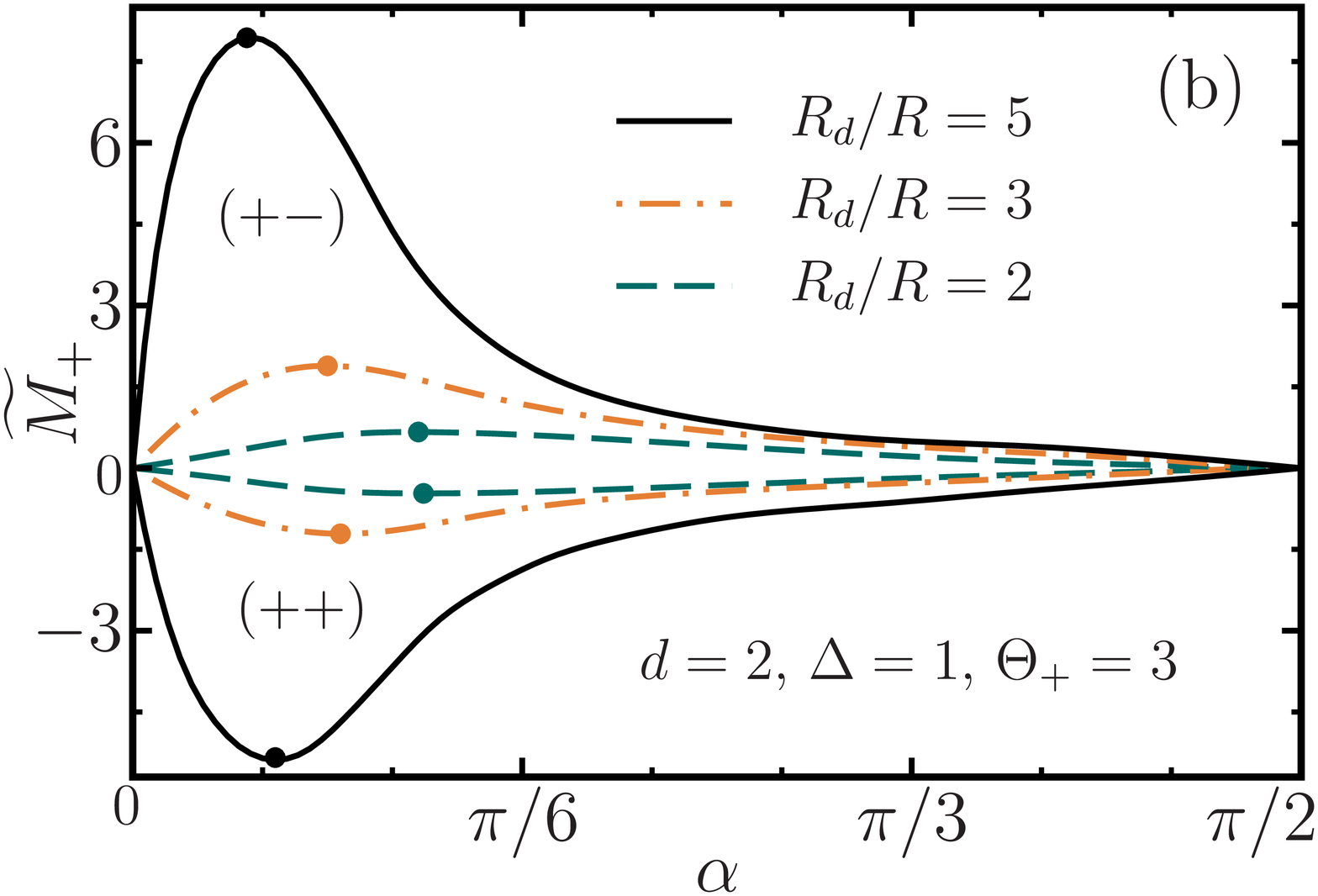}
\else
   \includegraphics*[width=0.4\textwidth]{agrs_d=3-torque}
   \includegraphics*[width=0.4\textwidth]{agrs_d=2-torque}
\fi
   \caption{(color online) The universal torque scaling function
   $\widetilde M_+ = M_+(\Theta_+, \Delta, \alpha; \,\delta)/ |
   K_+(\Theta_+=0, \Delta; \,\delta=1)|$ as a function of the angle
   $\alpha$ between the wall and the main axis of an elongated colloid
   (Fig.~\ref{fig:model}) for three values of the ratio $\delta=R_d/R$.
   In (a) $d=3$, $\Delta=L/R=2$, and $\Theta_+ = L/\xi_+ = 2$, while in
   (b) $d=2$, $\Delta=1$, and $\Theta_+ = 3$. The positive (negative)
   curves correspond to $+-$ ($++$) boundary conditions. $\widetilde M_+
   > 0$ ($\widetilde M_+<0$) implies that the torque acts as to
   increase (decrease) $\alpha$. The torque is expressed in terms of the
   absolute value of the Casimir force for $++$ boundary conditions at
   the critical point for a ($d=3$)-sphere with radius $R=R_1= L/2$ in
   (a) and for a disk ($d=2$) with radius $R=L$ in (b). The extrema are
   marked by full dots; their positions shift to smaller values of
   $\alpha$ upon increasing $R_d/R$. All curves correspond to keeping
   temperature and the \emph{minimal distance} $L$ fixed upon varying
   the orientation, and the pivot is taken to be the point closest to
   the wall (denoted as a circle in Fig.~\ref{fig:model}).
   \label{fig:torque:angle}}
 \end{center}
\end{figure}

\begin{figure}[!ht]
 \begin{center}
\ifOnecolumn
   \includegraphics*[width=0.49\textwidth]{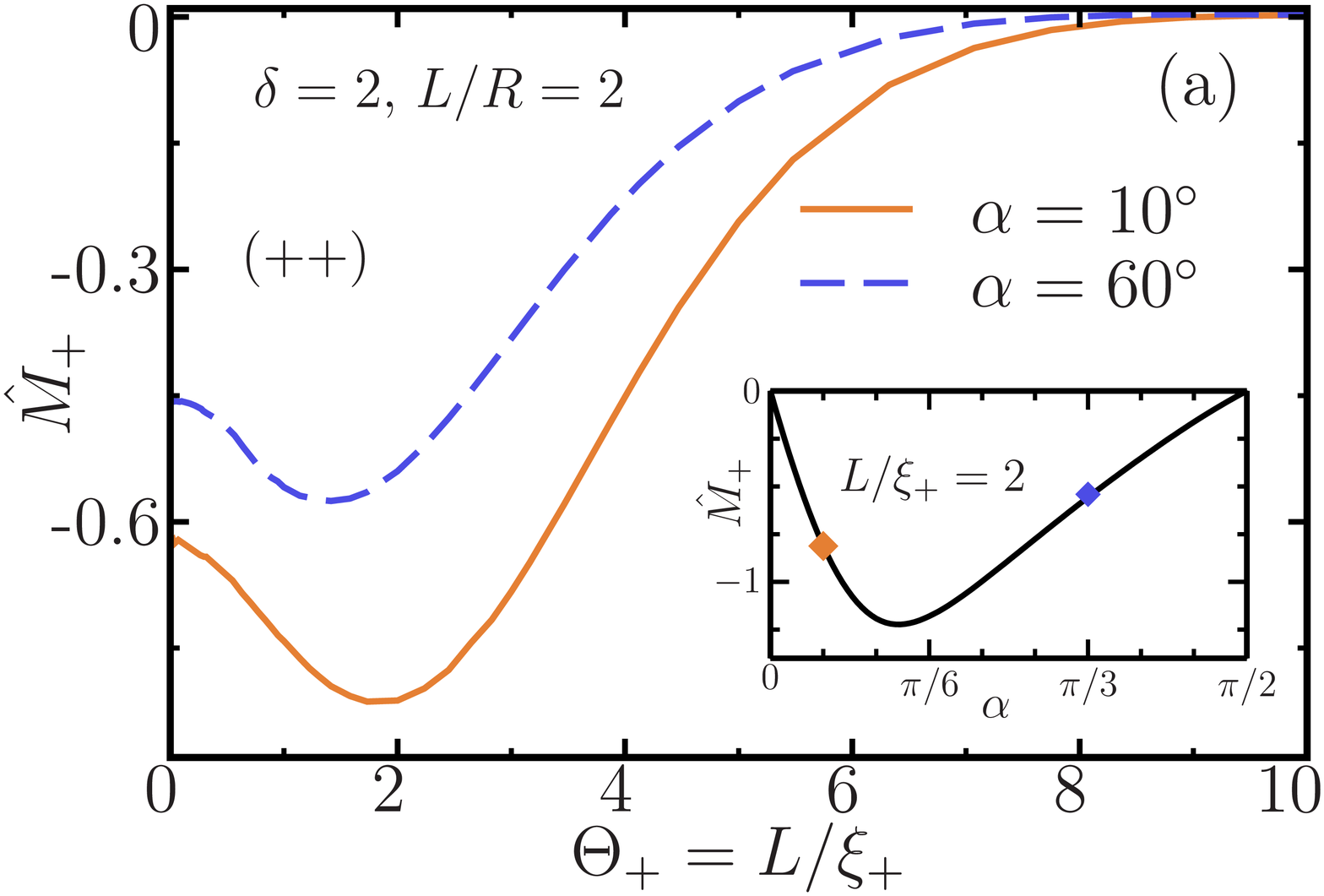}
   \includegraphics*[width=0.49\textwidth]{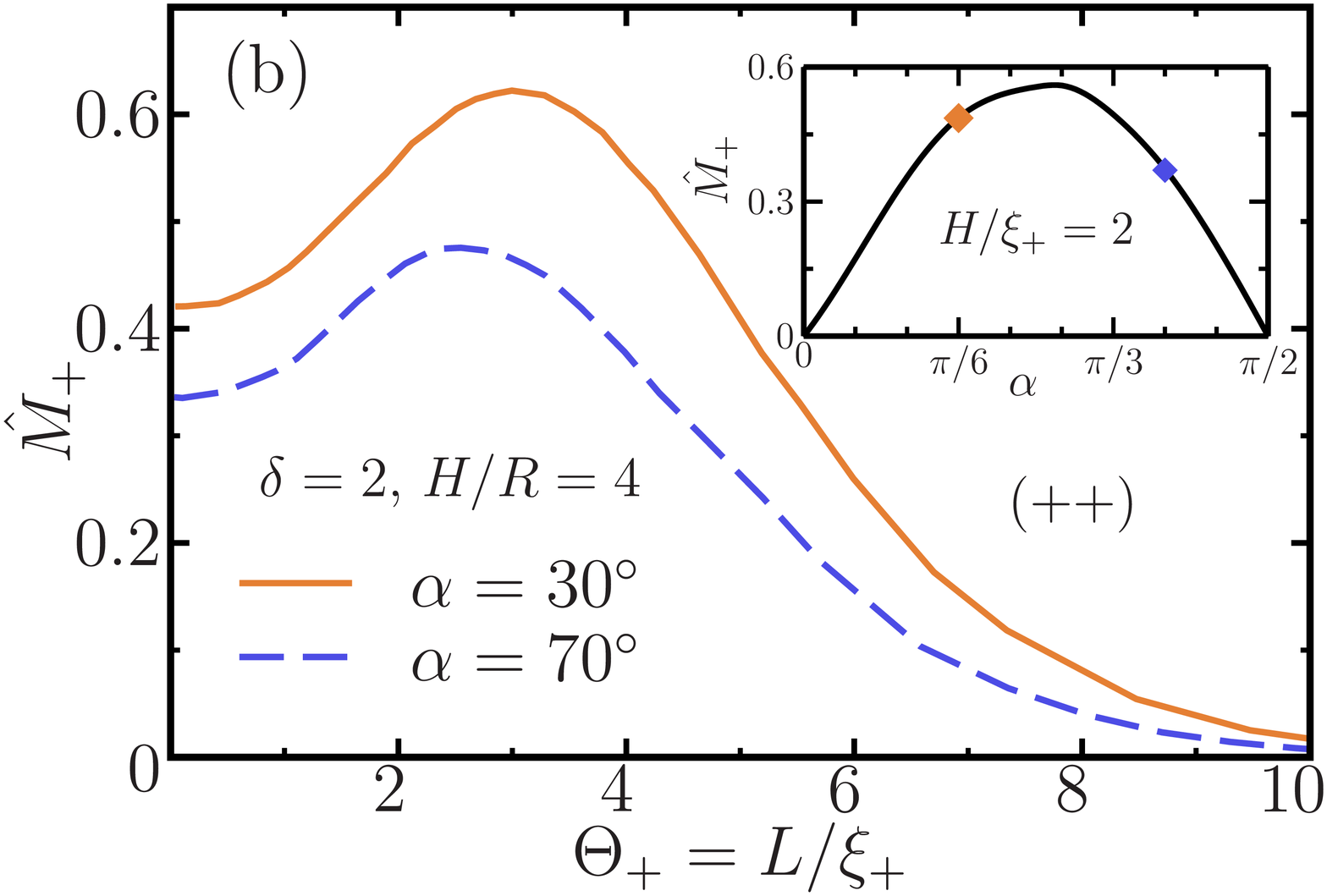}
\else
   \includegraphics*[width=0.4\textwidth]{agrs_torque_L}
   \includegraphics*[width=0.4\textwidth]{agrs_torque_H}
\fi
   \caption{ (color online)  The universal torque scaling function
   $\hat M_+ =  M_+(\Theta_+, \Delta, \alpha; \,\delta=2) /
   K^{(\parallel)}_{+}(\Theta_+=0)$ as a function of temperature for a
   spheroid (spheroido-cylinder ${\cal K}_{d=3}$ in $D>d$) with
   $\delta=R_d/R=2$ for $++$ boundary conditions and for two values of
   the angle $\alpha$ between the main axis of the spheroid and the wall
   (Fig.~\ref{fig:model}).  In (a) the \emph{surface-to-surface}
   distance $L$ is fixed with $\Delta=L/R=2$ and the torque is
   calculated with respect to the point of closest approach (denoted as
   a circle in Fig.~\ref{fig:model}). In (b) the distance $H$ between
   the wall and the \emph{center of the ellipsoid} is fixed with $H/R =
   4$ and the torque is calculated with respect to the center of the
   ellipsoid (denoted as a square in Fig.~\ref{fig:model}). The insets
   show $\hat M_+$ as a function of $\alpha$ for
   $L/\xi_+=2=\mathrm{const}$ in (a) and $H/\xi_+=2=\mathrm{const}$ in
   (b). The symbols in the insets correspond to the values of the angle
   $\alpha$ represented in the main plots. The torque is expressed in
   term of the absolute value of the Casimir force for $++$ boundary
   conditions at the critical point for the film geometry. In both (a)
   and (b) the torque is strongest at a distinct temperature above (and
   not at) $T_c$.
   \label{fig:torque:distance}}
 \end{center}
\end{figure}

The torque scaling function $M_+$ (see Eqs.~(\ref{eq:T}),
(\ref{eq:torque}), and (\ref{eq:momentum})) as a function of the angle
$\alpha$ (Fig.~\ref{fig:model}) between the main axis of an elongated
colloid and the wall is presented in Fig.~\ref{fig:torque:angle} for
both $++$ and $+-$ boundary conditions and for a fixed rescaled minimal
distance $L/\xi_+ = \mathrm{const}$ with the pivot point
$\boldsymbol{r}^{(1)}$ (see Eq.~(\ref{eq:momentum})) taken to be the
point closest to the wall (denoted as a circle in Fig.~\ref{fig:model}).
The scaling function is positive for $++$ and negative for $+-$ boundary
conditions. This means that in the $++$ case the configuration is
optimal (i.e., the free energy is lowest) if the colloid is parallel
($\alpha=0$) while in the $+-$ case if it is perpendicular to the wall
($\alpha = \pi/2$). As one may expect, the torque vanishes for $\alpha =
0$ and $\alpha = \pi/2$, which for $+-$ ($++$) boundary conditions
correspond to a  maximum (minimum) and a minimum (maximum) of the free
energy, respectively. Also as expected, the magnitude of the torque
increases upon increasing the ratio $R_d/R_1$ (for fixed $R_1/L$). 

As we have already mentioned, one can introduce different pivots with
respect to which the torque is exerted on the ellipsoid, such as the
point of closest approach to the wall (denoted as a circle in
Fig.~\ref{fig:model}) and the center of the ellipsoid (denoted as a
square in Fig.~\ref{fig:model}). The latter is more convenient to use if
the ellipsoid is far from the substrate. In this case the motion of the
ellipsoid can be described in terms of its center of mass (which we
assume to coincide with the geometrical center); accordingly the
orientational degrees of freedom of the ellipsoid should consistently be
described also with respect to the center of the ellipsoid. If the
ellipsoid is sufficiently close to the substrate, it is more convenient
to monitor the orientations of the ellipsoid with respect to the point
of closest approach to the wall. One reason is that in the region close
to the substrate not all orientations of the ellipsoid with respect to
its center of mass are allowed because the particle cannot penetrate the
substrate (c.f., Sec.~\ref{sec:comparison} and
Fig.~\ref{fig:orientations}(a)). One can also imagine that the ellipsoid
is trapped in a relatively shallow potential well of optical tweezers.
In this situation the rotation of the ellipsoidal particle occurs
naturally with respect to the point closest to the wall.

In Figs.~\ref{fig:torque:distance}(a) and \ref{fig:torque:distance}(b)
the torque scaling function $M_+$ with a suitable normalization is
plotted for these two choices of the pivot for a few values of the angle
$\alpha$ between the main axis of the ellipsoid and the wall and for
$++$ boundary conditions. In Fig.~\ref{fig:torque:distance}(a) the
torque is calculated with respect to the point closest to the wall, and
consistently we consider a constant minimal rescaled distance $L/\xi_+ =
\mathrm{const}$ for all angles. In Fig.~\ref{fig:torque:distance}(b) the
torque is calculated with respect to the center of the ellipsoid, so
that the rescaled distance between the center of the ellipsoid and the
wall is fixed, i.e., $H/\xi_+=\mathrm{const}$. Note that the scaling
function is negative in the former and positive in the latter case. This
means that if the surface-to-surface distance $L$ is kept fixed (for
instance, by optical tweezers or by the wall), the optimal configuration
of an ellipsoidal colloid is to be parallel to the wall ($\alpha = 0$).
However, if the ellipsoid rotates with respect to its center of mass,
the optimal configuration is to be perpendicular to the wall ($\alpha =
\pi/2$). In the $+-$ case the situation is reverse.

\section{Comparison with the quantum-electrodynamic Casimir interaction
and the  polymer induced depletion interaction}
\label{sec:comparison}

\begin{figure*}[!th]
  \begin{center}
\ifOnecolumn
   \includegraphics*[width=\textwidth]{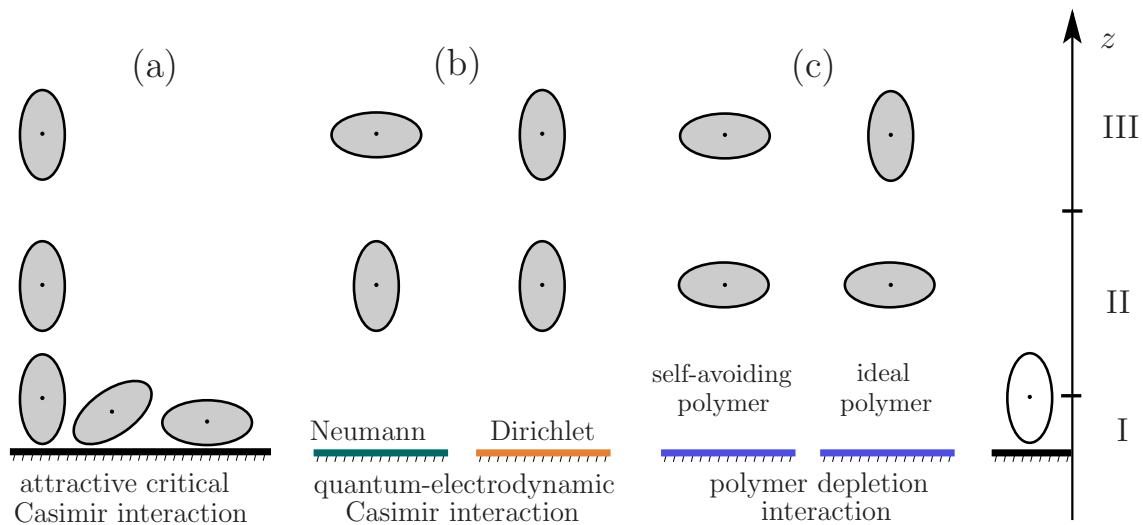}
\else
   \includegraphics*[width=0.85\textwidth]{figs_comparison2}
\fi
    \caption{(color online) Illustration of the preferred orientation of
    a prolate ellipsoid at a fixed  distance of its center from a planar
    wall in the $z$-direction. Only the projection onto the plane of the
    figure is shown. The centers of the ellipsoids are chosen to lie
    within the plane of the figure. (a) In the case of the attractive 
    critical Casimir interaction the ellipsoid is oriented perpendicular
    to  the wall in regions II and III, while it prefers to touch the
    wall in  region I because in that configuration the corresponding
    Casimir force is stronger than for other  orientations. (b) In the
    case of a scalar model of quantum-electrodynamics and for Neumann
    boundary conditions on the surface of the ellipsoid, a change  of
    the preferred orientation has been found for Neumann boundary
    conditions at the wall \cite{emig:08}. For Neumann boundary
    conditions on the particle surface and Dirichlet boundary conditions
    at the wall the energy is minimal if the ellipsoid  is perpendicular
    to the wall in both regions II and III. In the case of mixed
    Dirichlet and Neumann boundary conditions the 
    quantum-electrodynamic Casimir force is repulsive. (c) Using the
    small particle  operator expansion it has been shown that the
    attractive polymer induced depletion interaction leads to a change
    of the preferred orientation of the ellipsoid in the case of the
    ideal polymers, while the favorable orientation is the parallel one in
    both regions II and III in the case of self-avoiding polymers
    \cite{eisenriegler:03, eisenriegler:06:204903,
    eisenriegler:06:144912}. The preferred orientation of an ellipsoidal
    particle in region I has not yet been studied for the
    quantum-electrodynamic Casimir force or the polymer depletion
    induced interactions. The extension of region II in the $z$-direction is
    of the order of the length of the long axis of the ellipsoid in (b)
    and given by  the radius of gyration of the polymers in (c).}
  \label{fig:orientations}
  \end{center} 
\end{figure*}

The quantum-electrodynamic Casimir interaction \cite{rodriguez:08,
emig:08} and the polymer induced depletion  interaction
\cite{eisenriegler:03, eisenriegler:05, eisenriegler:06:204903,
eisenriegler:06:144912} lead to pronounced effects on the orientational
ordering of non-spherical particles, too. The schematic presentation in
Fig.~\ref{fig:orientations} illustrates the influence of these two
interactions and of the critical Casimir interaction on the orientation
of a prolate  ellipsoidal particle near a planar wall. For the purpose
of the discussion, the  region adjacent to the wall can be divided into
three sub-regions denoted as I, II, and  III (see
Fig.~\ref{fig:orientations}). In region I not all orientations of the
prolate ellipsoid are allowed because the particle cannot penetrate the
wall. In the case of an attractive critical Casimir interaction the
prolate ellipsoid reaches its most favorable  configuration of lying
parallel to the wall by tilting such that for a prescribed distance of
its center from the wall its optimum angle in region I is the one for
which it is in touch with the wall  (Fig.~\ref{fig:orientations}(a)).
The influence of both the quantum-electrodynamic Casimir interaction and
the polymer induced depletion interaction on the orientation of an
ellipsoid in region I has not yet been studied. For larger distances
from the wall, in regions II and  III, the critical Casimir torque
drives the prolate ellipsoid into an orientation  perpendicular to the
wall (Fig.~\ref{fig:orientations}(a)). In the case of the
quantum-electrodynamic Casimir interaction and of the polymer induced
depletion interaction the preferred orientation in region II is
perpendicular \cite{emig:08} (Fig.~\ref{fig:orientations}(b)) and
parallel \cite{eisenriegler:03, eisenriegler:06:204903,
eisenriegler:06:144912} (Fig.~\ref{fig:orientations}(c)) to the wall, 
respectively. Upon further increasing the distance from the wall a
change of the  preferred orientation has been found in the case of a
scalar model of quantum-electrodynamics with Neumann boundary conditions
both at the surface of the particle and at the wall \cite{emig:08}
(Fig.~\ref{fig:orientations}(b)) and in the case of ideal polymers
acting as depletion agents \cite{eisenriegler:03}
(Fig.~\ref{fig:orientations}(c)).

The polymer induced depletion interaction has been calculated 
\cite{eisenriegler:03, eisenriegler:06:204903,    
eisenriegler:06:144912} in the so-called protein limit in which the size
of  the ellipsoid is small compared to the polymer size characterized by
the radius of  gyration. By using the small particle operator expansion,
it has been shown that the extension of region II along the
$z$-direction is given by the radius of gyration  of the polymers acting
as depletion agents \cite{eisenriegler:03, eisenriegler:05,
eisenriegler:06:204903, eisenriegler:06:144912}. However, the presently
available non-spherical colloidal particles are larger than typical
polymers and monitoring small particles by optical techniques is very
difficult. Therefore, it would be rewarding to study theoretically and
experimentally the polymer induced depletion interaction beyond the
small particle limit. 

We emphasize that the preferred orientations of a prolate ellipsoid due
to the quantum-electrodynamic Casimir interaction
(Fig.~\ref{fig:orientations}(b)) have been obtained by considering a
\emph{scalar} model instead of the actual vectorial electromagnetism. In
the actual case of the electromagnetic field, which implies Dirichlet
boundary conditions, and a slightly deformed sphere, the preferred
orientation of a prolate ellipsoid is the one perpendicular to the wall
in both regions II and III \cite{emig:08}. Future work in  this area may
focus on the understanding of the influence of the
quantum-electrodynamic  Casimir interaction on ellipsoids of arbitrary
eccentricities and at small distances from the wall.

\section{Conclusions}
\label{sec:summary}

We have investigated the critical Casimir effect for single
non-spherical colloidal particles immersed in a fluid near its critical
point and exposed to a laterally homogeneous planar wall. For an
ellipsoidal colloidal particle the resulting critical Casimir force and
torque can be characterized by universal scaling functions $K_\pm$
(Eq.~(\ref{eq1})) and $M_\pm$ (Eq.~(\ref{eq:T})), respectively, which
depend on the dimensionless scaling variables $\Theta_\pm = L/\xi_\pm$,
$\Delta = L/R_1$, the angle $\alpha$ between the main axis of the
ellipsoid and the wall, and the ratios $\delta_{i-1}=R_i/R_1$, where
$R_1$ is the smallest semi-axis of the ellipsoid, $R_i$ ($i=2,\cdots,d$)
are $d-1$ remaining semi-axes, $L$ is the closest distance between the
surface of the ellipsoid and the wall (see Fig.~\ref{fig:model}), and
$\xi_\pm$ is the bulk correlation length above ($+$) and below ($-$) the
critical point. The scaling functions have been calculated within
mean--field theory, which represents the leading order term in a
systematic $\epsilon = 4-D$ expansion. The dependence of the scaling
functions $K_\pm$ on the scaling variable $\Theta_\pm$ exhibits
behaviors which are qualitatively similar to those of a spherical
colloidal particle (see Figs.~\ref{fig:force:distance_comp} and
\ref{fig:force:distance}). The strength of the force depends on the
orientation of the colloidal particle relative to the wall: for
elongated colloids it is stronger if the colloid is oriented parallel to
the wall at the same surface-to-surface distance $L$ (see
Figs.~\ref{fig:comparison2} and
\ref{fig:force:angle}). If ellipsoidal colloids are oriented
perpendicular to the wall, the force is stronger for shorter colloids
(see Fig.~\ref{fig:force:angle}). We note, however, that the latter
effect is due to the specific shape of the ellipsoid at its elongated
edge (see Subsec.~\ref{sec:mf:force}), while the former effect is more
general.

The sign of the universal torque scaling function $M_+$ depends on the
boundary conditions and on the pivot with respect to which the particle
rotates. Thus, if the pivot is chosen to be the point on the particle
surface closest to the wall (denoted as a circle in
Fig.~\ref{fig:model}), the scaling function is positive for equal ($++$) 
and negative for opposing ($+-$) boundary conditions at the wall and at
the particle surface (see Figs.~\ref{fig:torque:angle} and
Fig.~\ref{fig:torque:distance}(a)). This means that \emph{at the same
closest distance} $L$ between the surfaces of the colloid and the wall
an elongated particle tends to orient itself parallel to the wall for
equal boundary conditions and perpendicular to the wall for opposing
boundary conditions. We expect this situation to be realized if a
particle is close to the wall or, for instance, if it is trapped by
optical tweezers. An opposite effect is observed if the center of the
particle is kept fixed (see Fig.~\ref{fig:torque:distance}(b)). In this
case an elongated colloid prefers an orientation perpendicular to the
wall for equal boundary conditions and parallel to the wall for opposing
boundary conditions. It is worthwhile to note that this conclusion
agrees qualitatively with the asymptotic results obtained from the
small-particle operator expansion \cite{eisenriegler:04} (see also
Fig.~\ref{fig:orientations}(a)).

\acknowledgements
S.~K. and L.~H. gratefully acknowledge support by grant HA 2935/4-1 of
the Deutsche Forschungsgemeinschaft.

\bibliographystyle{apsrev}

\end{document}